\documentclass[sigconf]{acmart}

\setcopyright{none}
\copyrightyear{2026}
\acmYear{2026}
\acmDOI{10.1145/3772318.3790968}

\acmConference[CHI '26]{the 2026 CHI Conference on Human Factors in Computing Systems}{April 13--17, 2026}{Barcelona, Spain}

\acmISBN{979-8-4007-2278-3/26/04}

\makeatletter
\providecommand{\protected@file@percent}{\%}
\makeatother

\usepackage[noabbrev,capitalise,nameinlink]{cleveref}
\usepackage{algorithm}
\usepackage{algorithmic}
\usepackage{amsmath}
\usepackage{enumitem}
\usepackage{xspace}
\usepackage{xcolor}
\usepackage{multirow}
\usepackage{makecell}

\begin{document}

\title{No Pixel Left Behind: Filling Gaps in Anime Colorization}

\newcommand{\systemName}{\textit{GapFill}\xspace}

\author{Masahiro Kono}
\orcid{0009-0002-3432-8541}
\affiliation{%
  \institution{The University of Tokyo}
  \city{Tokyo}
  \country{Japan}
}
\email{marckono2825@g.ecc.u-tokyo.ac.jp}

\author{Akinobu Maejima}
\orcid{0000-0002-8005-9218}
\affiliation{%
  \institution{OLM Digital, Inc.}
  \institution{IMAGICA GROUP Inc.}
  \city{Tokyo}
  \country{Japan}
}
\email{akinobu.maejima@olm.co.jp}

\author{Yuki Koyama}
\orcid{0000-0002-3978-1444}
\affiliation{%
  \institution{The University of Tokyo}
  \city{Tokyo}
  \country{Japan}
}
\email{koyama@pe.t.u-tokyo.ac.jp}

\author{Yotam Sechayk}
\orcid{0009-0002-5286-0080}
\affiliation{%
  \institution{The University of Tokyo}
  \city{Tokyo}
  \country{Japan}
}
\email{sechayk-yotam@g.ecc.u-tokyo.ac.jp}

\author{Takeo Igarashi}
\orcid{0000-0002-5495-6441}
\affiliation{%
  \institution{The University of Tokyo}
  \city{Tokyo}
  \country{Japan}
}
\email{takeo@acm.org}

\renewcommand{\shortauthors}{Kono, et al.}

\begin{abstract}
  
Animation production workflows often involve digital colorization of line art, where small unpainted regions (``gaps'') frequently occur and remain an underexplored challenge.
We conducted a formative study in Japanese animation (anime) pipelines and found that while the paint bucket tool is widely used for base coloring, tiny enclosed areas are frequently overlooked, resulting in time-consuming manual detection and filling.
We introduce \systemName, a tool grounded in professional practices that reduces the effort of gap detection, zooming, and color selection.
Our deep-learning method suggests appropriate fill colors by referencing surrounding regions, leveraging the flat-color nature of anime-style images.
In a user study with $13$ professional colorists, our system improved performance and usability in gap-filling tasks over conventional methods. The study also suggested that prediction accuracy alone is not the primary factor for usability, that appropriate colors can be contextually ambiguous, and that \systemName can complement existing tools depending on users' trust in new AI-powered assistance.

\end{abstract}



\begin{teaserfigure}
  \includegraphics[width=\textwidth]{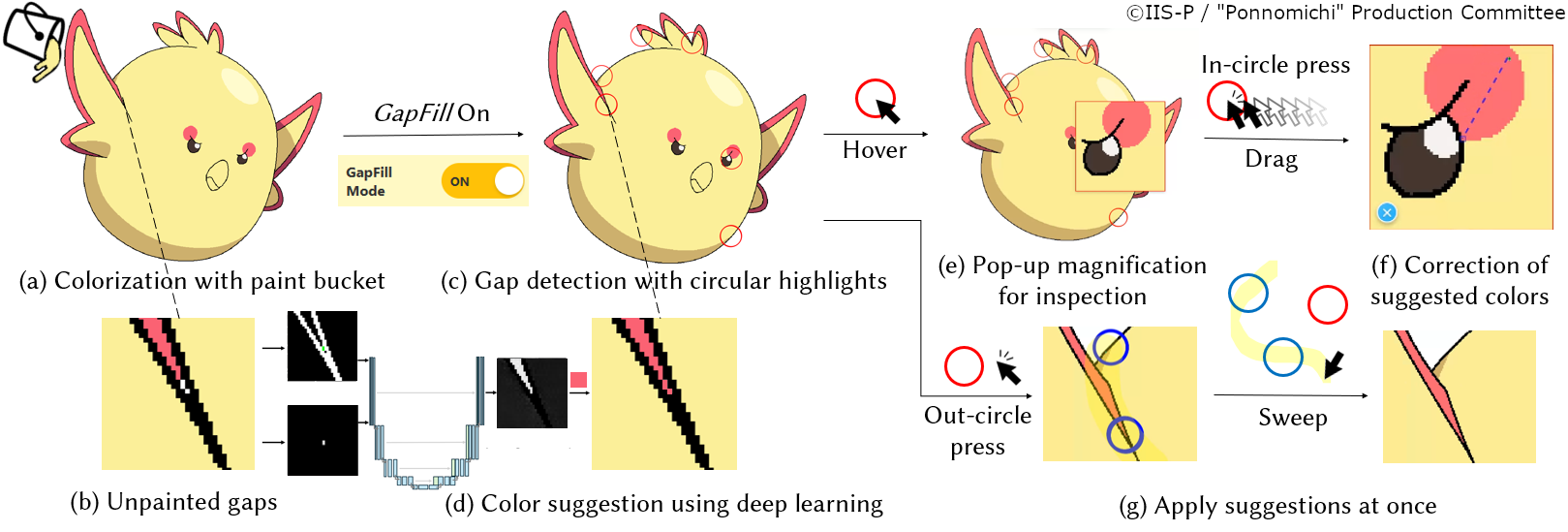}
\caption[Teaser figure]{
  \systemName assists professional anime colorists in addressing small unpainted gaps.
  (a) Colorization with the \textit{paint bucket} often (b) leaves small enclosed regions (``gaps'') unpainted. 
  When \systemName is activated, (c) gap detection with circular highlights is triggered, and (d) these gaps are temporarily filled with suggested colors using our domain-specific deep learning method.
  (e) Hovering over a highlight shows a magnified view, allowing inspection without zooming. 
  (f) Dragging within a highlight activates a color-pick mode for correcting the suggestion.
  (g) Users can sweep across correct suggestions to apply them at once.
}
  \Description{A diagram illustrating \systemName workflow using a yellow bird character as an example. The figure is divided into steps (a) through (g).
Left side: (a) shows a paint bucket tool filling the character, which results in (b) small unpainted white gaps between the fill and line art.
Center: (c) shows the \systemName toggle being turned on, which triggers gap detection marked by red circular highlights on the character. (d) depicts a domain-specific U-shaped neural network architecture (U-Net) processing the image to predict colors for these gaps.
Right side: Three interaction techniques are shown. (e) Hovering the cursor over a highlight reveals a magnified pop-up window for inspection. (f) Clicking and dragging inside a highlight allows the user to manually correct the suggested color. (g) Clicking outside the highlights and sweeping the cursor across multiple circles applies the suggested colors in a batch.}
  \label{fig:teaser}
\end{teaserfigure}

\maketitle

\section{Introduction}
\label{sec:introduction}

Japanese animation (\textit{anime}), deeply rooted in Japanese pop culture, has emerged as a globally recognized form of media art.
The anime industry is notable for its cultural influence and economic significance, supported by a rapidly expanding international market~\cite{AnimeIndustryReport} and a vast global fan base~\cite{MyAnimeList}.
Despite a high volume of broadcasts (over 200 titles per year~\cite{BroadcastStats}), the \textit{colorization} process, referring to filling flat colors into each region of hand-drawn line art for every frame, remains largely manual. This reflects the legacy of traditional 2D anime production practices, where works were painted on sheets known as~\textit{cels} \cite{otsuka_2022_animemade}. This process transitioned to digital production in the mid-1990s~\cite{ichikohji2013influence} and is now primarily supported by digital painting software such as Clip Studio Paint (CSP)~\cite{CLIPSTUDIO}.

\begin{figure}[t]
  \centering
  \includegraphics[width=0.95\linewidth]{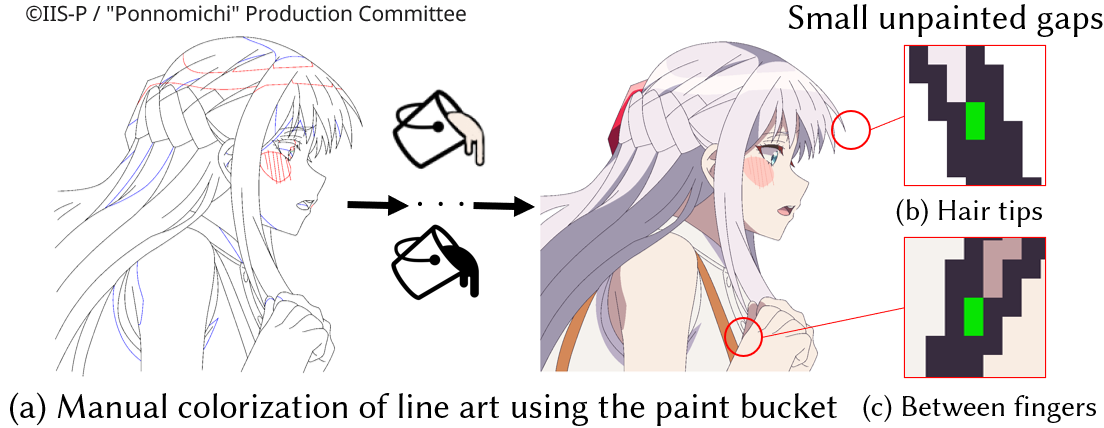}
    \caption[Flood fill and gaps]{
Colorization with the \textit{paint bucket} often leaves small unpainted ``gaps'' (green), tedious to detect and fill.
    }
    \Description{An illustration showing the occurrence of unpainted gaps during the anime colorization process.
(a) displays the workflow where line art of a female anime character is colored using a paint bucket tool. While the character appears mostly colored, red circles highlight specific small problematic areas.
(b) shows a magnified pixel-level view of one such area at hair tips, revealing a small, narrow region highlighted in green. This represents a ``gap'' where the paint bucket failed to fill.
(c) shows another magnified view between fingers, similarly displaying a unpainted gap trapped between the outlines.
The figure visually demonstrates how standard filling tools leave small artifacts in sharp corners.}
  \label{fig:flood_fill_gaps}
\end{figure}

Although digital colorization is central to anime production pipelines, the process itself has received little academic attention.
To address this gap and gain a deep understanding of professional workflows, we conducted a two-stage formative study with 20 colorists, including interviews with 4 from a commercial studio. Results revealed the typical anime colorization workflow: starting from binary (usually non-anti-aliased) line art composed of contours and color-coded guides for shadows and highlights, colorists sequentially apply base colors to segmented regions while referencing the model sheet.
In this process, the \textit{paint bucket} (\textit{flood fill}) tool, which employs region-growing algorithms to fill enclosed areas bounded by lines~\cite{fourey2018fast}, was found to be widely used~(\cref{fig:flood_fill_gaps}a).

Additionally, our formative study substantiated a practical yet underexplored challenge that existing tools fail to adequately address: specialized support for filling small unpainted enclosed regions, hereafter referred to as \textbf{``gaps''}.
These gaps typically result from unintentional line intersections or isolated minor regions within the line art.
They frequently appear in sharp-angled areas such as hair tips~(\cref{fig:flood_fill_gaps}b) or narrow spaces between fingers~(\cref{fig:flood_fill_gaps}c), and are often difficult to detect visually.
Unlike hobbyist illustration, professional colorization adheres to strict standards where even a single unpainted pixel necessitates a costly retake. Consequently, detecting and filling such gaps remains a time-consuming burden, emphasizing the need for production-oriented tools.

We designed and developed \textbf{\systemName} (\cref{fig:teaser}), a specialized tool for gap-filling that integrates seamlessly into existing professional pipelines.
The system enables automatic gap detection with circular highlights, along with temporary filling based on color suggestions using our domain-specific deep learning method. By hovering over a highlight, the corresponding region can be magnified for quick inspection without zooming in. By dragging within a highlight, a color-picker-like interface is activated for correcting the color suggestion. For cases where suggestions appear accurate, users can apply them at once either by sweeping across multiple highlights or by using a one-click fill button.
The key design goal was to reduce the repetitive burden of manually detecting, zooming, and selecting colors for such gaps, while fitting naturally into conventional workflows and maintaining user control over AI assistance.

To enable automatic color prediction, we propose a domain-specific deep learning method formulated as a localized inference problem. Our approach leverages the characteristics of anime images, typically composed of flat-color regions. Our method predicts plausible colors from the surrounding context. Instead of directly regressing color values, our model estimates a spatial likelihood map that identifies the neighboring region most likely to share the same color as the target region. This region-to-region correspondence enables the model to infer colors indirectly and robustly, suiting the flat and discrete nature of anime-style coloring.

The performance and usability of \systemName were evaluated through a user study with 13 professional colorists. Participants completed two structured tasks comparing \systemName against conventional tools: (1) coloring real-world anime line art from scratch, and (2) detecting and filling unpainted gaps in partially colored images. These tasks were designed to reflect both the full colorization workflow and the final checking process. Task performance was measured by completion time and the number of overlooked gaps, while perceived usability was assessed through surveys, semi-structured interviews, and feature-level analyses. The results demonstrated significant improvements, particularly in the second task. Our findings suggest that \systemName has the potential to complement existing tools by leveraging their strengths, and also offer insights into professionals’ trust in new AI-powered assistance. Moreover, our color prediction method achieved an accuracy of $81.68\%$ on an unseen dataset, and its outputs were subjectively rated as valuable in production contexts. Notably, the study indicated that usability was shaped not only by prediction accuracy, partly because the appropriate color can be ambiguous depending on the context; participants also valued clear visual aids and the controllability of AI suggestions.

To summarize, our contributions are:
\begin{itemize}
    \item A formative study filling the gap between anime production and research, capturing real-world colorization workflows and identifying the overlooked challenge of unpainted gaps.

    \item \systemName, a specialized tool for colorists to fill these gaps that integrates seamlessly into professional pipelines.

    \item A deep learning method that predicts colors for unpainted regions by leveraging the flat-color nature of anime-style images and inferring local context.

    \item An evaluation with 13 professionals showing that \systemName improves task efficiency, with findings indicating that usability is not driven by prediction accuracy alone, that appropriate colors can be contextually ambiguous, and that the adoption of AI-powered assistance depends on users' trust.
\end{itemize}
Our code is available at \url{https://marc2825.github.io/GapFill}.

\section{Related Work}
\label{sec:relatedWork}

\subsection{AI-Powered Creativity Support Tools for Digital Painting}

Various \textit{Creativity Support Tools} (CSTs)~\cite{shneiderman2007creativity} have been proposed for digital painting in HCI and CG.
\citet{frich2018hci,frich2019mapping} map the landscape of creativity research via systematic reviews and characterize CSTs.
LazyBrush~\cite{sykora2009lazybrush} colors imprecise drawings via energy minimization, while KISSColor~\cite{dong2025kisscolor} infers closed regions in vector sketches via kinetic stroke stretching along winding-number fields. 
FlatMagic~\cite{yan2022flatmagic} supports professional comic artists in flat coloring using neural rendering and intermediate representations, whereas Painting with Bob~\cite{benedetti2014painting} targets novices, prioritizing ease of use.
Color Portraits~\cite{jalal2015color} characterizes key color manipulation activities via human-centered design and inspires novel interaction tools beyond traditional pickers. 
Colorbo~\cite{kim2022colorbo} supports interactive mandala coloring via AI-generated suggestions.
\citet{bao2019scribble} proposed a scribble-based tool for diffusion-curve-based vector colorization. 
AniFaceDrawing~\cite{huang2023anifacedrawing} leverages StyleGAN with a two-stage training strategy to transform incomplete sketches into high-quality anime portraits.
Collectively, these tools demonstrate the potential of AI-driven methods in digital painting, aligning with our study.

As AI-powered tools become common, HCI research has turned toward the user perspective, emphasizing issues of trust, acceptability, and explainability in human–AI collaboration.
Co-drawing studies show creators prefer controllable assistance with explanations~\cite{oh2018lead}, and that AI collaboration yields quality comparable to human-only work~\cite{fan2019collabdraw}. \citet{pei2024human} further demonstrate that AI involvement shapes cognitive load and creative efficacy. Beyond drawing, two-way communication enhances engagement and perceived reliability~\cite{rezwana2022understanding}, though artists report tensions regarding authenticity~\cite{bird2024artists}. Adoption studies note that trust predicts acceptance better than technical features~\cite{xu2023everyone}, while explainability and control remain essential~\cite{liapis2022need}.
Building on this, we empirically examine creative professionals' perceptions of new AI-powered tools.

\subsection{Automatic Colorization for Line Art}

Many studies have explored automatic colorization for line art used in anime and manga drawings.
Classical approaches include color propagation methods that are aware of patterns and textures~\cite{qu2006manga}, bipartite matching based on regions across frames~\cite{kanamori2012region}, and methods that perform matching between a graph constructed from a reference image and a target image~\cite{sato2014reference}.
Early learning-based approaches include PaintsChainer~\cite{pfnet_paintschainer}, which employs CNNs for automatic colorization and Style2Paints~\cite{zhang2017style}, a fully automatic feed-forward model used for applying specific painting styles to anime sketches.
Advancements in deep learning techniques have given rise to various models, such as a two-stage GAN framework that mimics human workflows~\cite{zhang2018two}, a U-Net-based model that skips low-confidence regions~\cite{ishii2020confidence}, and flat color prediction for comics using ResNet-based classifiers and Transformer models~\cite{verduyn2024towards}.
Recently, diffusion-based models tailored for this domain have achieved high performance~\cite{cao2024animediffusion}, followed by several subsequent approaches~\cite{yan2025image, liu2025manganinja}.
However, most of these methods rely on fully colorized reference images or user hints, limiting flexibility.

Recent studies have also explored learning-based colorization under limited examples~\cite{maejima2024continual} and a robust matching approach that formulates region correspondence as a set of inclusion relationships~\cite{dai2024learning}.
However, these methods often struggle with small or intricate regions even in production settings where reference frames are available, limiting their practical applicability to the proposed problem setting.
For small-region colorization, \citet{akita2020colorization} proposed a method to fill empty pupils in line art, but its scope remains limited. A related task of our setting is image inpainting, where many deep learning approaches have been proposed~\cite{zhang2023image}, but these mainly target natural photos with continuous tones and gradients, unlike anime-style drawings with flat and discrete regions.
In the domain of semi-automated, user-guided colorization, \citet{zhang2021user} proposed a system that interprets user scribbles to control color propagation interactively, \citet{ci2018user} introduced a conditional GAN model conditioned on both line drawings and user-provided strokes, and \citet{zou2019language} further extended control through language-based inputs.
Nevertheless, these frameworks are not fully suited to real-world anime production workflows.

\subsection{Understanding Anime Production}

Academic research on anime production has traditionally focused on socio-cultural perspectives in fields such as anthropology, sociology, and media studies.
For example, \citet{condry2013soul} attributes anime's global success to social energy across industry and fans, while \citet{morisawa_2015_managing} shows how creative authority often outweighs management in anime studio hierarchies, and \citet{mihara_2020_coming} expands the analytical lens by foregrounding anime's business personnel and advocating a business anthropology approach.
Psychological approaches also examine expressive techniques in anime, such as \citet{yokota_2019_psychology} on achieving emotional impact with limited frames.

\citet{kato2025anime} positioned anime as an emerging topic in HCI, integrating technical, cultural, and industrial perspectives to support production and foster an international research community.
Similarly, \citet{ichikohji2013influence} examined the integration of digital technologies in studios to analyze their impact.
For production support, Griffith~\cite{kato2024griffith} specializes in anime storyboarding (e-conte) deriving general findings for CSTs, and AnimAgents~\cite{wang2025animagents} is a collaborative system that streamlines animation pre-production by orchestrating AI tools.
In computer vision and graphics, challenges in supporting anime pipelines have driven research on super-resolution for final outputs~\cite{wang2024apisr}, generating in-betweens from key frames~\cite{xing2024tooncrafter}, and non-photorealistic rendering methods to replicate anime aesthetics, such as \citet{todo2024practical}'s 3D style transfer pipeline. 
Comprehensive survey papers on AI applications in cel animation~\cite{tang2025generative, rai2025sketchanimationstateoftheartreport} further consolidate this growing body of computer science-based work.
Taken together, these studies illustrate a convergence of cultural and technical perspectives on anime production, opening opportunities both to deepen scholarly understanding and to develop practical tools tailored to production contexts.

\begin{figure*}[t]
  \centering
  \includegraphics[width=\linewidth]{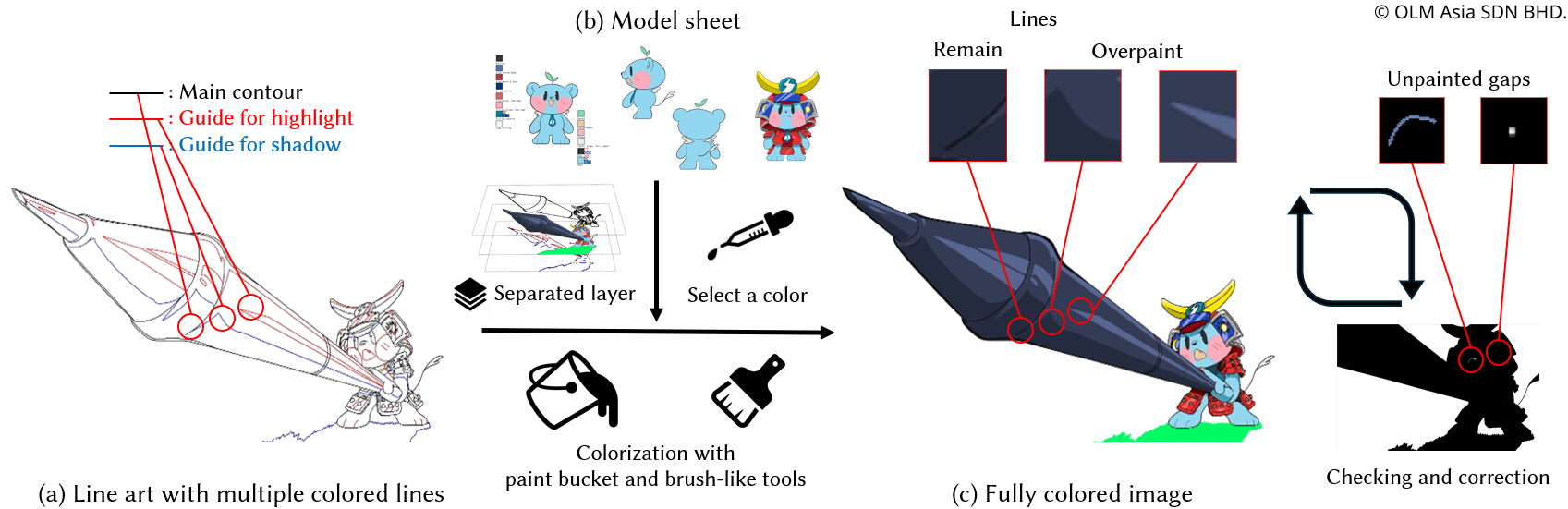}
  \caption[Colorization workflow]{
    Typical anime colorization workflow: (a) preparing line art with main contours and highlight/shadow guides, (b) selecting colors from model sheets and applying them on separated layers, and (c) ultimately producing the fully colored image.
  }
\Description{A flow diagram illustrating the standard industrial workflow for anime colorization, divided into three stages labeled (a) through (c) and a final quality check.
Stage (a) ``Line art with multiple colored lines'' displays a drawing where different line colors serve specific functions: black lines represent main contours, red lines are guides for highlights, and blue lines are guides for shadows.
Stage (b) shows the process of selecting colors from a character ``Model sheet'' and applying them using paint bucket or brush-like tools on separated layers.
Stage (c) ``Fully colored image'' presents the final output. Zoomed-in callouts labeled ``Lines'' demonstrate that main black contours ``Remain'' visible, whereas guide lines (such as red highlight guides) are ``Overpainted'' and disappear.
On the far right, a cyclic step titled ``Checking and correction'' illustrates the detection of ``Unpainted gaps'' (shown as white artifacts on a black silhouette), indicating an iterative process to fix these errors.}
  \label{fig:colorization_workflow}
\end{figure*}

\newpage

\section{Understanding Anime Colorization Workflow: A Formative Study}
\label{sec:background}

The professional anime production pipeline relies on a specialized division of labor. In this workflow, \textit{``colorization''} is a critical component of the \textit{``shiage''} (finishing) process~\cite{kato2024griffith, otsuka_2022_animemade}, where colorists apply colors to clean binary line drawings provided by in-between animators from the preceding \textit{``douga''} stage. The colored frames are then submitted to the subsequent \textit{``composition''} stage, where compositors adjust and integrate them with other assets, such as background art and 3DCG, to complete the scene. Since the colored image is passed directly to this phase, strict adherence to color specifications is required to ensure the visual quality of the final output. Even minor coloring errors can be treated as defects that triggers a retake, increasing both labor costs and delays.

Despite its importance, the specific practices and challenges of the colorization process remain understudied.
To address this gap, we conducted a formative study with professional colorists in a commercial studio.
Guided by the principle that practical adoption depends on integration with existing workflows~\cite{chilana2015user}, we aimed to understand production realities and derive design principles for tools suitable for real-world deployment. In doing so, we answer calls for CSTs research to better support expert practitioners and their practices~\cite{frich2019mapping, frich2018hci}.

\subsection{Anime Colorization in Practice}
\subsubsection{Procedure of The Formative Study}

To capture both the overall process and hidden challenges in real workflows, we conducted a two-stage formative study: a broad-scale questionnaire (S1) followed by in-depth semi-structured interviews (S2)~\cite{lazar2017research}.

\begin{itemize}
\item \textbf{S1 (Questionnaire):} We distributed an online survey to 20 professional colorists (R1--R20) via a manager at an anime studio. The survey covered their experience, tool usage, and perceptions of the colorization workflow and its challenges. Participants had $1$ to $9$ years of experience ($M = 4.6$, $SD = 2.3$). For the 7-point Likert items, ratings of 5 or above were treated as positive (results are reported as $M$ and $SD$).

\item \textbf{S2 (Interview):} To gain deeper insights into the findings from S1, we conducted 30-minute semi-structured interviews with 4 experienced professionals (I1--I4) recruited from the same studio (experience range: $3$--$5$ years). All participants provided consent for recording and transcription.

\end{itemize}
We additionally obtained a screen recording of a professional colorization process (V1) from the anime studio to visually examine the workflow and complement the self-reported data. This material compensated for the lack of mandatory screen sharing in S2, which was made optional considering the confidentiality of the assets.

\subsubsection{Grounding Systematic Understanding of Anime Colorization}

\begin{figure*}[t]
  \centering
  \includegraphics[width=0.9\linewidth]{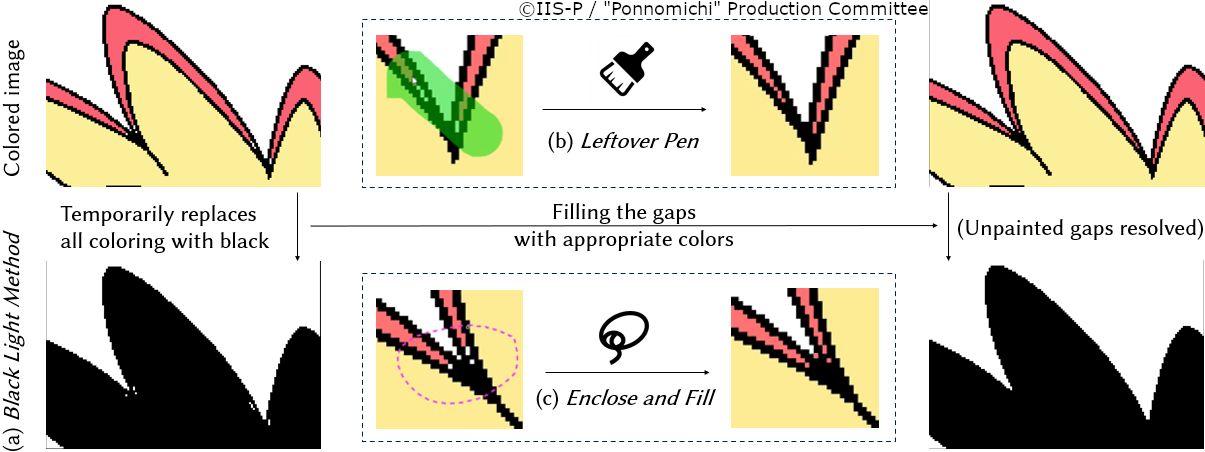}
  \caption[Black light method and other conventional tools]{
    Conventional methods for addressing unpainted small gaps. (a) \textit{Black Light Method}: Temporarily replace all colors with black in a single click, making gaps appear as white dots for easy detection. (b) \textit{Leftover Pen}: Fill all gaps along the stroke with the specified color. (c) \textit{Enclose and Fill}: Fill all gaps within the enclosed area using a lasso-like tool.
  }
\Description{A diagram illustrating three conventional methods for fixing unpainted gaps in digital illustration, labeled (a) through (c).
(a) ``Black Light Method'': The left side compares a ``Colored image'' with a ``Black Light'' view. In the colored image, small white gaps are hard to see against light colors. In the Black Light view, all painted areas are temporarily rendered as solid black, making the unpainted white gaps stand out clearly as high-contrast bright spots.
(b) ``Leftover Pen'': A technique where a user draws a broad stroke (shown in green) over a gap. The tool automatically detects the unpainted area within the stroke and fills it with the user-specified color.
(c) ``Enclose and Fill'': A technique where a user draws a lasso-like loop (shown as a pink dotted line) around a cluster of gaps. All unpainted areas inside the loop are filled instantly with the user-specified color.
The right side shows the final result where all gaps are resolved, resulting in a clean image in both color and silhouette views.}
  \label{fig:blacklight}
\end{figure*}

\paragraph{Tools and Environment}
All $24$ participants (S1 \& S2) used CSP~\cite{CLIPSTUDIO} for colorization.
Input devices in S1 consisted of pen tablets ($80\%$) and pen displays ($70\%$); all participants relied on at least one of these stylus interfaces, while $35\%$ additionally used a mouse and keyboard as auxiliary inputs.
This diversity underscores the importance of designing an interface that accommodates a range of input modalities.
Regarding software tools, the \textit{Paint Bucket} was the most frequently used ($90\%$), followed by brush tools ($50\%$) including the \textit{Leftover Pen} (\cref{fig:blacklight}b), AI-based auto-coloring tools ($40\%$), and lasso-like tools ($20\%$) such as \textit{Enclose and Fill} (\cref{fig:blacklight}c).

\paragraph{Standard Colorization Workflow}
According to demonstrations (I3, V1) and additional self-reports, the standard colorization process proceeds as follows.
The process begins with clean line art on separate layers, where main contours are drawn in black, highlight guides in red, and shadow guides in blue\footnote{Guide color conventions vary across studios; colors like green may be used.} (\cref{fig:colorization_workflow}a).
Colors are selected for each region based on the model sheet (\cref{fig:colorization_workflow}b), and applied to specific layers using tools such as the Paint Bucket and brush.
Finally, the highlight and shadow guides are painted over, producing a fully colored image (\cref{fig:colorization_workflow}c).
According to I1 and I4, the time required per frame varies significantly based on complexity, ranging from under a minute to $15$--$30$ minutes.
To improve efficiency, I1 described a batch-processing strategy: \textit{``I don't actually change the color in between. So, it's a bit faster compared to if you do one whole thing,''} implying that they color the same specific parts across multiple frames before switching colors, rather than completing frames one by one.

\paragraph{Limitations of Current AI Tools}
We also examined the studio's internal AI-based auto-coloring tool~\cite{maejima2024continual}.
According to I1, this tool transfers colors from adjacent frames, achieving $70$--$80\%$ accuracy for minor movements.
However, it frequently misrecognizes regions in dynamic scenes, leading I1 to remark, \textit{``most of the time we have to check on its work.''}
Consequently, professionals often avoid using it to prevent redundancy; as I2 stated, \textit{``in order not to waste my time, I’m rather to do manually because [...] do the painting twice,''} highlighting that \textbf{human control} is currently more reliable than AI automation.
Nevertheless, I2 expressed openness to future adoption if performance improves: \textit{``I think it should be a convenient tool [...] If there is a very good accuracy I would use.''}
This suggests that while AI tools are available, their usability may depend on accuracy.

\subsubsection{A Practical Challenge in Colorization: Small Unpainted Gaps}

We first asked participants an open-ended question in S1: \textit{``Do you encounter any common issues when coloring line art?''}
Participants highlighted a range of difficulties, including distinguishing similar colors (R1, R4, R19), human errors such as choosing incorrect colors (R5, R8, R13, R18), and managing complex layer structures (R6, R9, R17).
However, the most frequently cited issue was \textbf{small unpainted gaps}, voluntarily mentioned by $11$ out of $20$ participants (\textit{``small pixel unreachable when using bucket tool''} (R7), \textit{``filling too many small pixels''} (R20)).
Identifying this as a common bottleneck, we focused our subsequent inquiry on the specific workflows and perception regarding gap-filling.

\paragraph{Current Gap Detection and Filling Method}
We examined how such gaps are addressed based on demonstrations (I3, V1) and additional self-reports.
Professionals typically employ a visual check known as the \textit{``Black Light Method''} (\cref{fig:blacklight}a): temporarily replacing all colorings with black via a shortcut to make transparent gaps stand out as bright dots.
However, the method for filling these gaps varies.
I3 uses a specialized brush, the \textit{Leftover Pen}~\cite{CLIPSTUDIO} (\cref{fig:blacklight}b), which fills unpainted areas along a stroke.
In contrast, I1 uses a lasso-like \textit{Enclose and Fill} tool~\cite{CLIPSTUDIO} (\cref{fig:blacklight}c), which fills all gaps within the enclosed area, while the artist in V1 relied on the standard Paint Bucket.
Despite these variations, all methods require manual \textbf{detection}, \textbf{zooming}, and \textbf{color selection and filling}.

\paragraph{Workflow Strategies and Burden}
The burden of addressing gaps is substantial.
I3 noted that they detect gaps by \textit{``zooming in very close and moving the canvas bit by bit,''} a process that can take up to several minutes per frame in the worst case for complex drawings.
This was corroborated by S1, where $60\%$ of participants agreed with \textit{``Do you find addressing unpainted gaps to be time-consuming?''} ($M=5.0$, $SD=1.8$), confirming that gap-filling is a repetitive manual task.
Regarding timing, two distinct strategies emerged: I3 prefers a batch-processing approach, filling all gaps in a single pass at the end of coloring to minimize tool switching, whereas I4 fixes gaps immediately upon noticing them. I4 also emphasized the importance of \textbf{workflow consistency} of new tools, stating, \textit{``if they want to change into another program we have to learn it all over again.''}
Reflecting this diversity, an ideal tool and its evaluation should go beyond mere final inspection and be designed to \textbf{fit seamlessly into individual professionals’ practices}.

\paragraph{Frequency and Significance}
The quantitative results further validated the significance of this issue.
When asked \textit{``Do you often need to address unpainted gaps?''}, $65\%$ of participants in S1 responded positively ($M=4.8$, $SD=1.7$). I2 noted that such gaps appear in almost every image, particularly at thin, sharp ends such as hair tips and strands, at complex line intersections, and in fine details.
Crucially, $85\%$ agreed with the statement \textit{``Do you think deciding the appropriate color for unpainted gaps is important for completing an animation?''} ($M=6.0$, $SD=1.2$).
I1 explained the visual impact: \textit{``if the gaps exist [...] it will be transparent in the background,''} noting that dedicated fans \textit{``will look for any kind of blemish.''}
These findings indicate that while physically minute at a glance, unpainted gaps have a disproportionately large impact on visual quality.

\begin{figure*}[t]
  \centering
  \includegraphics[width=\linewidth]{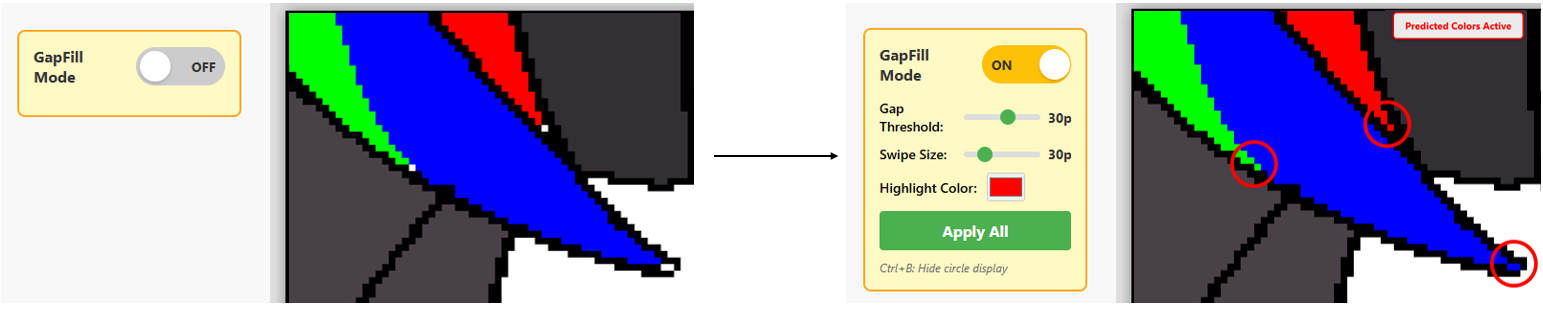}
  \caption[Overview of UI]{
    Overview of \systemName. When activated, the system automatically detects unpainted gaps, highlights them with circles, and temporarily fills them with suggested colors using a domain-specific deep learning method.
  }
    \Description{A side-by-side comparison showing the user interface and canvas before and after activating \systemName.
    Left (Inactive state): The ``\systemName Mode'' toggle is set to ``OFF''. The canvas shows a zoomed-in section of an illustration where small white artifacts (unpainted gaps) are visible at the sharp tips of red, blue and green shaded areas.
    Right (Active state): The ``\systemName Mode'' toggle is ``ON'', expanding the panel to show sliders for ``Gap Threshold'' and ``Swipe Size'', and an ``Apply All'' button. On the canvas, the unpainted gaps have been automatically detected and highlighted them with red circles (highlight colors can be changed via ``Highlight Color'' button). Then \systemName temporarily fills them with suggested colors using a domain-specific deep learning method that seems to match their surroundings.}
  \label{fig:overview_ui}
\end{figure*}

\paragraph{Decision Logic for Gap Filling}
Finally, we investigated how artists determine which color to use when filling a gap.
In S1, $90\%$ reported referencing surrounding colors, while $75\%$ selected official references such as model sheets, and $45\%$ checked adjacent frames.
I1 noted referring to surrounding elements and tones, relying on intuition when choices are ambiguous. I3 described first consulting model sheets to understand the subject, then determining colors by referencing adjacent frames and local context. These findings indicate that \textbf{local context} such as neighboring colors ultimately plays a central role in deciding which color to use for gap filling.

\subsection{Design Principles for \systemName}
Based on the findings, we identified conventional practices and production needs. Building on these insights, we derived the following design principles for \systemName, a tool tailored to \textbf{professional anime colorists} for addressing \textbf{small unpainted gaps}:

\begin{itemize}
    \item Reduce the manual workload in the repetitive cycle of \textbf{detection}, \textbf{zooming}, and \textbf{color selection and filling}.
    
    \item Fill gaps using \textbf{local context} such as neighboring colors, reflecting artists' reliance on local cues.
    
    \item Promote practical adoption~\cite{chilana2015user} via \textbf{intuitive, familiar interactions} that bridge conventional workflows and \textbf{support diverse, user-specific use cases}.
    
    \item Provide a \textit{Creativity Support Tool} that prioritizes \textbf{human controllability} rather than pursuing full automation, aligning with the findings of~\citet{roy2019automation}.
\end{itemize}

\section{Design and Implementation of \systemName}
\label{sec:system_overview}

\subsection{User Interface}
We developed the user interface (UI) shown in~\cref{fig:overview_ui}. The interface is activated on-demand via a toggle button to accommodate user-specific painting practices and provides five key functions: automatic detection of unpainted gaps with highlighting, deep learning-based color suggestions, a pop-up magnification for inspection, a color-pick-like operation for correcting color suggestions, and the application of suggested colors via a sweep-like interaction complemented by an apply-all button.

\subsubsection{Unpainted Gap Detector with Circular Highlights}
As shown in \cref{fig:teaser}c, the system automatically detects unpainted gaps and highlights them with circles around each detected region. A gap is defined as any enclosed, unpainted (transparent) region on the active coloring layer, with a pixel count below a user-adjustable threshold. Such boundaries may also be formed in combination with other layers, including line art and guide layers. Grid-based traversal algorithms such as BFS are used to identify such enclosed regions. It is also possible to employ methods like trapped-ball segmentation~\cite{zhang2009vectorizing, allen2024fast} to more strictly estimate regions by accounting for line discontinuities. This feature is designed to reduce the manual effort of detecting unpainted gaps.

\subsubsection{Automatic Color Suggestion for Filling Unpainted Gaps}
When \systemName is activated, each detected unpainted gap is temporarily overlaid with a suggested fill color using a deep learning–based prediction method (\cref{sec:system_overview:technique}). This feature aims to reduce the manual effort of selecting colors for filling unpainted gaps.

\subsubsection{Hover-Activated Pop-up Magnification for Quick Inspection}
As shown in \cref{fig:teaser}e, when the cursor is hovered over a highlight, a pop-up magnification (a fixed $5\times$ zoom independent of the canvas scale) is displayed; this is a feature inspired by Shift~\cite{vogel2007shift}. This magnification provides a localized magnified view centered on the detected unpainted gap. The user can quickly inspect the surrounding region that is temporarily filled with suggested colors, without manually zooming in the canvas. A hollow translucent marker at its center indicates the position of the detected gap. This feature is designed to reduce the manual effort associated with frequent zooming operations.

\subsubsection{In-Circle Color-pick for Correcting AI Suggested Colors}
We adopt a human-in-the-loop approach to complement occasional AI prediction errors. As shown in \cref{fig:teaser}f, this feature allows users to directly correct suggested colors for unpainted gaps.
When the pop-up magnification is visible (i.e., when the cursor is inside the circle), a color selection mode can be activated by
initiating a drag action. This mode behaves like a color picker: the pixel color under the cursor dynamically replaces the fill color of the corresponding unpainted gap. To clarify the substitution target, a dotted line connects the cursor to the center of the unpainted region when this mode is active. The pixel color at the release point of dragging becomes the final assigned color, resolving the unpainted state and removing the highlight. 
This feature enables users to correct suggested fill colors without zooming in and resolve a small number of mispredicted gaps via a familiar color-picker interaction, while adhering to the system requirement that emphasizes attention to local information during color selection.

\begin{figure*}[t]
  \centering
  \includegraphics[width=0.9\linewidth]{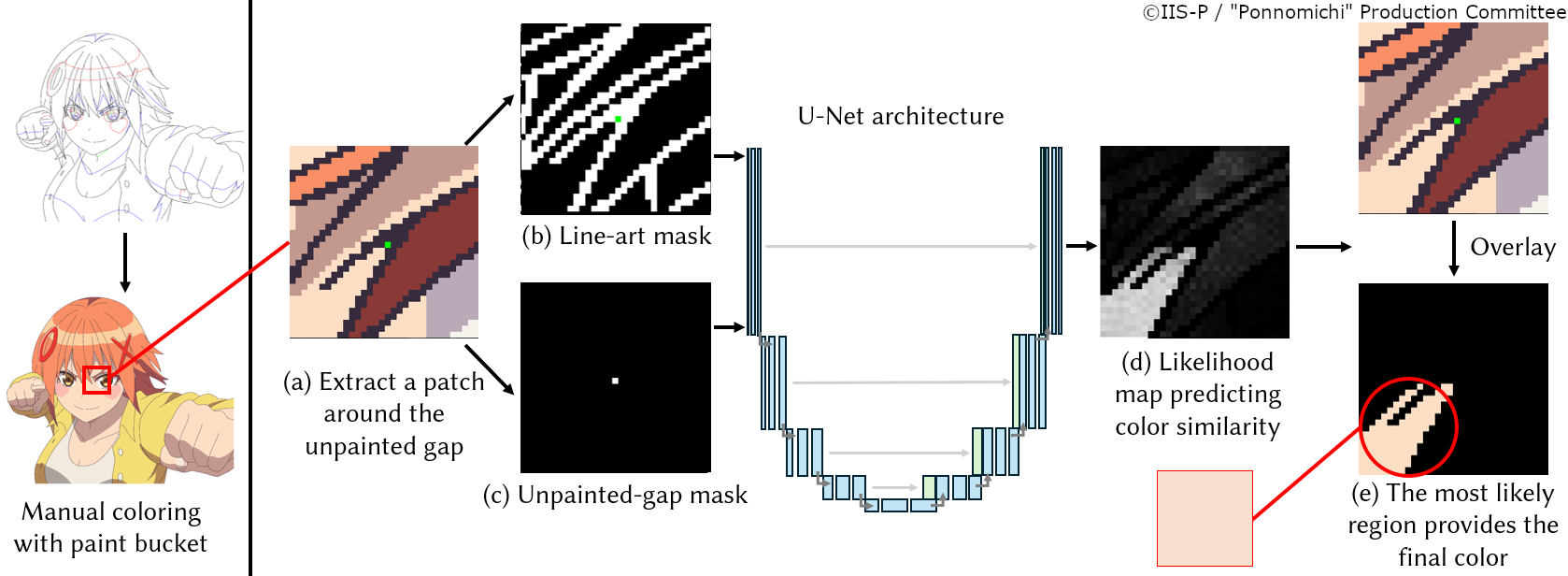}
  \caption[Color prediction method]{
    The color prediction method for \systemName.
    (a) First, we extract a patch around the target gap as a context for inference. (b) A \textit{line-art mask} and (c) an \textit{unpainted-gap mask} are input to the U-Net, which outputs (d) a likelihood map showing how likely each pixel matches the target color. (e) The final color is then chosen from the region with the highest likelihood.
  }
\Description{A schematic diagram illustrating our deep-learning pipeline for color prediction, flowing from left to right.
On the far left, an anime character under ``Manual coloring'' is shown, with a red box highlighting a specific unpainted gap.
(a) Shows the ``Extract a patch around the unpainted gap'' step, displaying a zoomed-in square image, which is treated as a local patch, around the gap containing surrounding colors and lines.
(b) and (c) show the inputs for the network: a ``Line-art mask'' (white lines on black background) and an ``Unpainted-gap mask'' (a small white area indicating the target pixels).
These are fed into a U-shaped ``U-Net architecture''.
(d) Shows the output ``Likelihood map'', a grayscale image where brighter pixels indicate a higher probability of being the correct color source.
(e) Shows the final step where the system identifies the ``most likely region that provides the final (output) color'' from the map while referring to the surrounding colors, and then samples that color (a skin tone) to fill the gap.}
  \label{fig:method_colorprediction}
\end{figure*}

\subsubsection{Out-Circle Sweep-to-Apply and Apply-All Button}
As shown in \cref{fig:teaser}g, when the suggested colors seem reasonable for the unpainted gaps, users can apply them in batches via a sweep-like interaction. When the cursor is outside a circular highlight and a drag action begins, the system enters the following mode. When dragging, all circles crossed by the translucent stroke are treated as selected and their corresponding unpainted gaps are marked for confirmation. Upon release, all selected gaps are simultaneously filled with their suggested colors, thereby resolving the unpainted regions and removing their highlights. This interaction was inspired by the painting metaphor for manipulating large sets of toggle switches~\cite{baudisch1998don}, which showed that paint-like gestures can make interaction more efficient. An \textit{Apply-All} button is provided to fill all unpainted gaps in a single click, offering an alternative option to accommodate diverse user preferences. These features enable users to efficiently adopt AI-driven color suggestions while preserving direct user manipulation via a familiar brush-like interaction.

\begin{figure}[t]
  \centering
  \includegraphics[width=\linewidth]{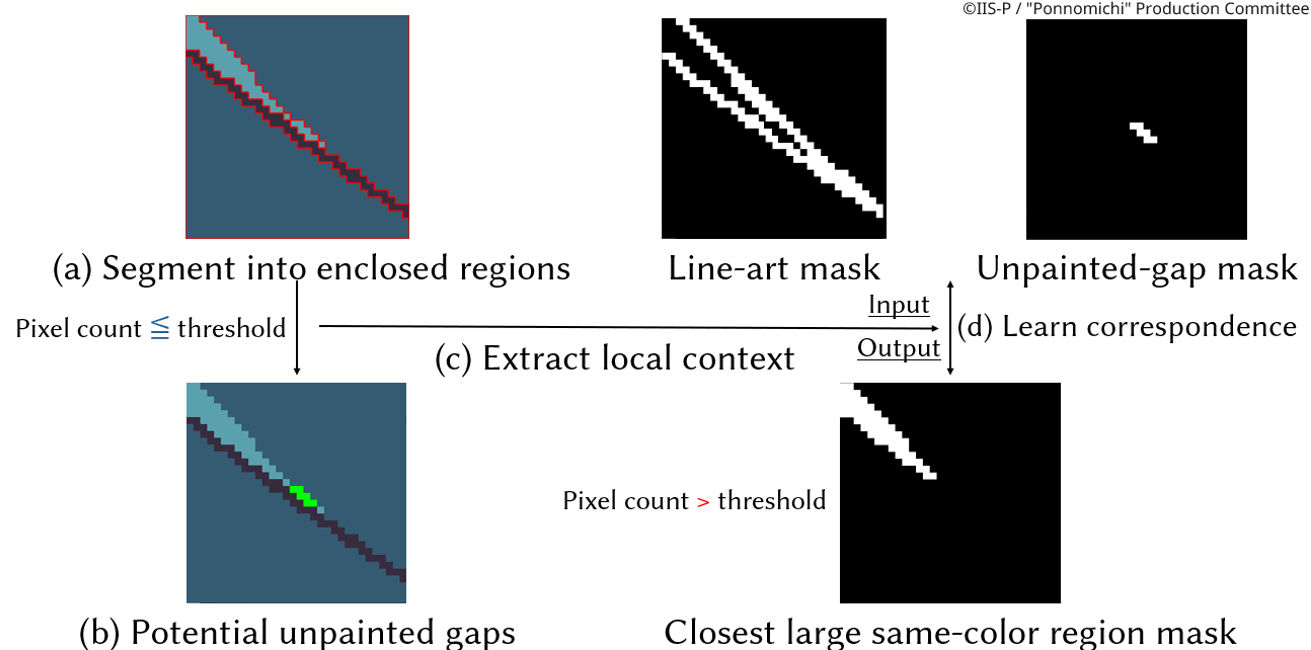}
    \caption[Training method]{Synthetic training dataset generation pipeline. (a) Region segmentation. (b) Identification of unpainted gaps below a pixel threshold. (c) Local context extraction. (d) Learning the mapping between input masks (\textit{line-art, unpainted-gap}) and output mask (\textit{closest large same-color region}).
  }
\Description{A flow diagram illustrating the pipeline for generating synthetic training data.
Steps (a) and (b) show the selection process: An image is ``Segmented into enclosed regions''. Small regions with a pixel count below a threshold are identified as ``Potential unpainted gaps'' (highlighted in green), simulating unpainted artifacts.
Step (c) ``Extract local context'' prepares the data for the model.
Step (d) ``Learn correspondence'' defines the machine learning task. It maps two ``Input'' images---a ``Line-art mask'' and the ``Unpainted-gap mask''---to a single ``Output'' image: the ``Closest large same-color region mask'' (a region with pixel count above the threshold). This teaches the model to associate small gaps with their surrounding large color regions, from which the appropriate fill color is derived.}
  \label{fig:training}
\end{figure}

\subsection{Method for Automatic Color Prediction}
\label{sec:system_overview:technique}

\subsubsection{Color Prediction via Region Correspondence}

Since the occurrence of unpainted gaps is independent of specific colors, our approach does not directly regress the colors themselves, but instead indirectly predicts them through the correspondence between regions. This enables us to construct a robust model that can predict flat, gradient-free colors, typical in anime-style images. To this end, we design a compact deep learning model based on U-Net~\cite{ronneberger2015u}, which is highly effective for generating segmentation masks and capturing features of neighboring regions, making it well-suited as a backend for interactive user interfaces. \cref{fig:method_colorprediction} illustrates the overall prediction framework.
The model takes a two-channel binary input: (\cref{fig:method_colorprediction}b) a \textit{line-art mask} where line pixels are set to $1$; and (\cref{fig:method_colorprediction}c) an \textit{unpainted-gap mask} where the target unpainted region is encoded as $1$.
The model then predicts a spatial likelihood map indicating the probability that each pixel within the patch shares the same color as the target area (\cref{fig:method_colorprediction}d). Finally, the suggested color is determined by selecting the color from the painted region with the highest average predicted likelihood (\cref{fig:method_colorprediction}e).

\subsubsection{Creating Synthetic Training Dataset}

Our formative study revealed that colorists rely on local context; our analyses of a professional anime image dataset (Appendix A) confirmed this, observing that small regions often share colors with neighbors. Grounded on these observations, we constructed a synthetic training dataset by applying BFS-based fill operations to line drawings to segment them into enclosed regions (\cref{fig:training}a). We then defined potential unpainted gaps as regions with pixel counts below a threshold of 10 (\cref{fig:training}b). For each such region extracted from the complete set of professional anime episodes ($1,807,977$ targets in total), we generated a $32 \times 32$ image patch centered on it as local context (\cref{fig:training}c) and applied data augmentation techniques such as rotation and flipping. Using these patches, we trained a model to map the \textit{line-art mask} and the \textit{unpainted-gap mask} (inputs) to the \textit{closest large} (i.e. above the pixel threshold) \textit{same-color region} mask  (output), computed from the ground-truth colored image (\cref{fig:training}d).

\begin{figure*}[t]
  \centering
  \includegraphics[width=0.9\linewidth]{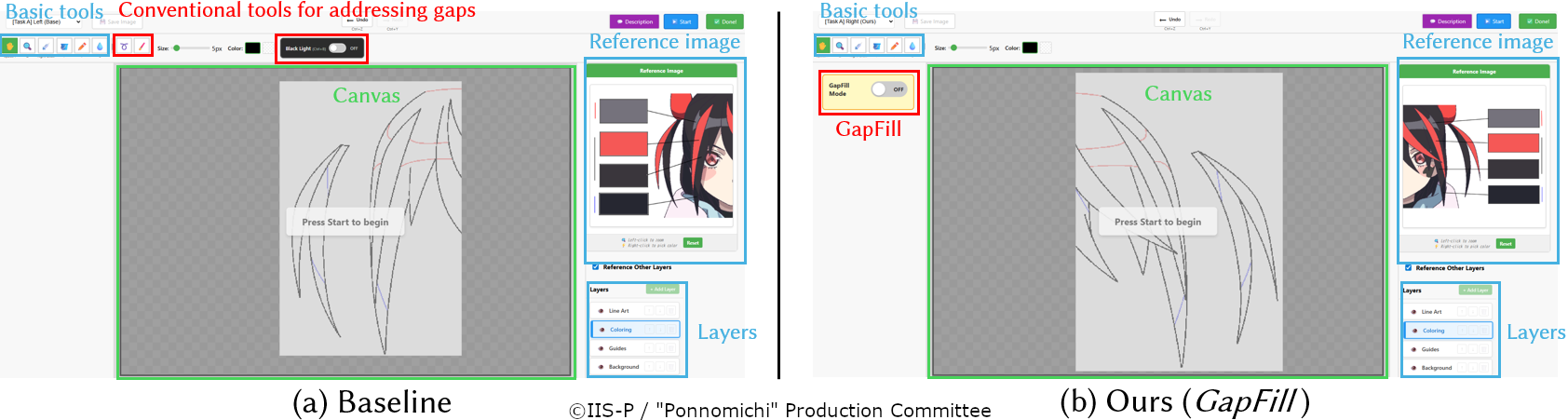}
    \caption[Overview of custom painting software and task A]{Overview of the custom painting software used in the user study, showing images for Task A. In addition to the basic features indicated by the blue and green boxes, (a) the \textit{Baseline} UI provides three commonly used tools for addressing gaps, whereas (b) the \textit{Ours} UI introduces \systemName, with the difference highlighted in red. %
    }
\Description{A side-by-side comparison of the custom painting software interfaces used in the user study, labeled (a) ``Baseline'' and (b) ``Ours''.
Both interfaces share a common layout with standard components outlined in blue and green: ``Basic tools'' at the top left, a large central ``Canvas'' displaying line art of anime hair, a ``Reference image'' modeled after the ``model sheet'' on the right, and a ``Layers'' panel at the bottom right.
The key difference lies in the red-highlighted toolbar area:
In (a) ``Baseline'', this area contains icons for conventional gap-filling tools (\textit{Leftover Pen} and \textit{Enclose and Fill}) and a \textit{``Black Light Method''} toggle switch.
In (b) ``Ours'', these tools are replaced by a single ``\systemName Mode'' toggle switch.
This visualizes how the experimental condition changed only the specific tools for addressing gaps while keeping the rest of the colorization environment identical.}
  \label{fig:userstudy_ui_taska}
\end{figure*}

\section{User Study with Professionals}
\label{sec:evaluation}

We conducted a user study with professional colorists to evaluate \systemName. The evaluation focused on task performance and perceived usability, with emphasis on gap detection and filling, reflecting insights from the formative study (\cref{sec:background}). Furthermore, qualitative feedback was collected on participants' experiences with the AI-powered tool and their impressions of individual features, regarding its potential for practical adoption in real environments.
This study was approved by our institution's ethics review board.

\subsection{Methodology}

\subsubsection{Participants}
Professional colorists were recruited from the same anime studio as in our formative study and $14$ individuals (P1--P14) participated voluntarily. They received compensation equivalent to their standard hourly wage and provided consent for both screen recording and transcription of their remarks. P13 was excluded from the subsequent analyses due to technical issues.

According to the pre-study questionnaire, all participants were adults under the age of $34$ and reported using CSP~\cite{CLIPSTUDIO} for their daily colorization. Their professional experience ranged from $1$ to $10$ years ($M = 3.6$, $SD = 2.6$). \cref{tab:participants} summarizes the basic demographics, including primary devices and main tools. %

\begin{table}[t]
\small
\centering
\caption[Participant demographics]{Participant demographics (P1--P14). Professional Experience in years; Devices: PT = \textit{Pen Tablet}, PD = \textit{Pen Display}, MK = \textit{Mouse and Keyboard}; Main Tools: PB = \textit{Paint Bucket}, LP = \textit{Leftover Pen}, EF = \textit{Enclose and Fill}, BL = \textit{Black Light}.
}
\Description{A table summarizing the demographics of 13 participants (labeled P1 through P14).
It consists of four columns: ``ID'', ``Professional Experience'' (in years), ``Devices'', and ``Main Tools''.
Professional experience ranges from 1 year (P5, P11, P14) to 10 years (P4).
The ``Devices'' column indicates hardware usage, abbreviating ``Pen Tablet'' as PT, ``Pen Display'' as PD, and ``Mouse and Keyboard'' as MK.
The ``Main Tools'' column lists the tools habitually used by each participant for addressing gaps: PB (``Paint Bucket''), LP (``Leftover Pen''), EF (``Enclose and Fill''), and BL (``Black Light'').
The data shows a diverse group of professionals using various combinations of hardware and conventional tools.}
\label{tab:participants}
\begin{tabular}{cccc}
\toprule
ID & Professional Experience & Devices & Main Tools \\
\midrule
P1 & 7 & PT, PD & PB, BL \\
P2 & 3 & PT, PD, MK & PB, LP, EF, BL \\
P3 & 2 & PT, PD & PB \\
P4 & 10 & PD, MK & LP, EF, BL \\
P5 & 1 & PD, MK & PB, LP \\
P6 & 2 & PT, PD, MK & PB, LP, BL \\
P7 & 2 & PT, MK & PB, BL \\
P8 & 3 & PT, MK & PB, LP, BL \\
P9 & 4 & PD, MK & PB, LP, BL \\
P10 & 5 & PD & PB, LP, BL \\
P11 & 1 & PD & EF \\
P12 & 5 & PD & PB, EF, BL \\
P14 & 1 & PT, MK & PB, LP, EF, BL \\
\bottomrule
\end{tabular}
\end{table}

\subsubsection{Procedure Overview}

\begin{figure}[t]
  \centering
  \includegraphics[width=\linewidth]{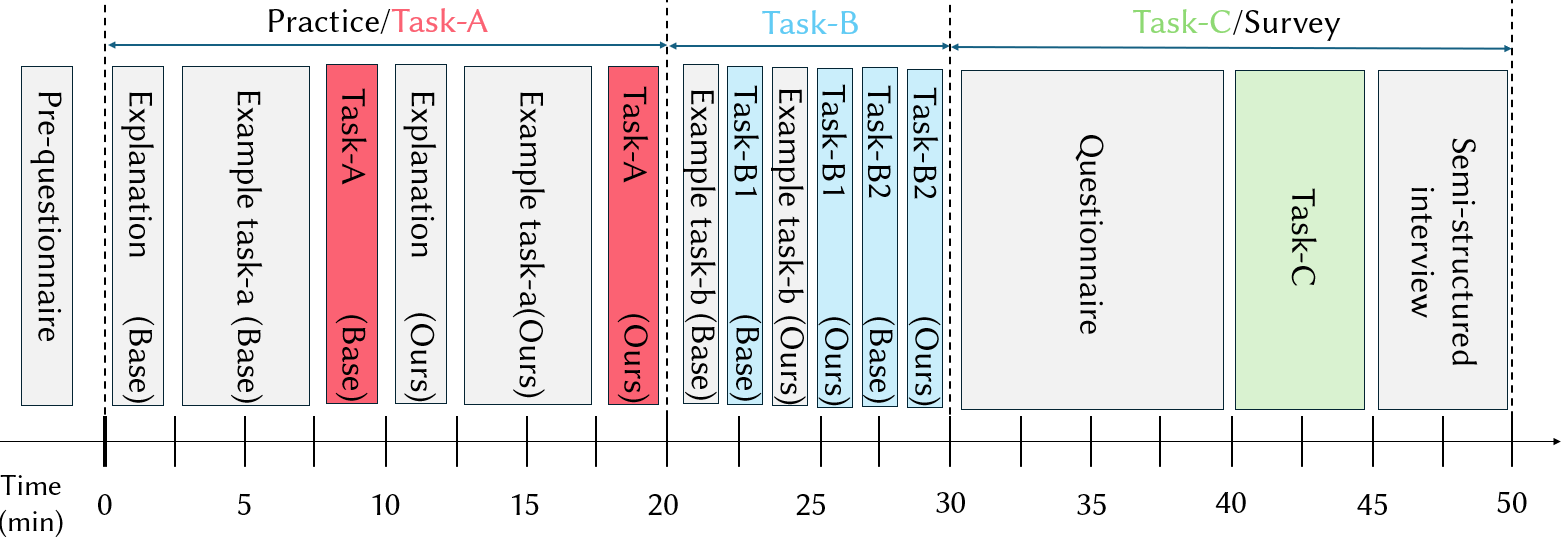}
    \caption[Overview of the user study procedure]{
      Overview of the user study procedure (example)
    }
\Description{A timeline diagram (example) illustrating the 50-minute procedure of the user study, divided into three main sessions labeled ``Practice/Task-A'', ``Task-B'', and ``Task-C/Survey''.
The process begins with a ``Pre-questionnaire''.
The first session ($0--20$ min) alternates between ``Base'' and ``Ours'' conditions, containing blocks for ``Explanation'', ``Example task'', and the main ``Task-A''.
The second session ($20--30$ min) follows a similar pattern for ``Task-B'', comparing ``Base'' and ``Ours'' across sub-tasks B1 and B2.
The final session ($30--50$ min) consists of a large block for ``Questionnaire'', followed by ``Task-C'', and concludes with a ``Semi-structured interview''.}
  \label{fig:userstudy_flow}
\end{figure}

The study was conducted online and a custom web-based painting software was used (\cref{fig:userstudy_ui_taska}). Each session, lasting $60$ minutes, was conducted in a one-on-one format, to enable direct interaction with each participant. Participants' shared screens and audio\footnote{Audio was transcribed via Google Gemini, with errors corrected for readability.} were recorded for subsequent analyses.

\begin{figure*}[t]
  \centering
  \includegraphics[width=0.9\linewidth]{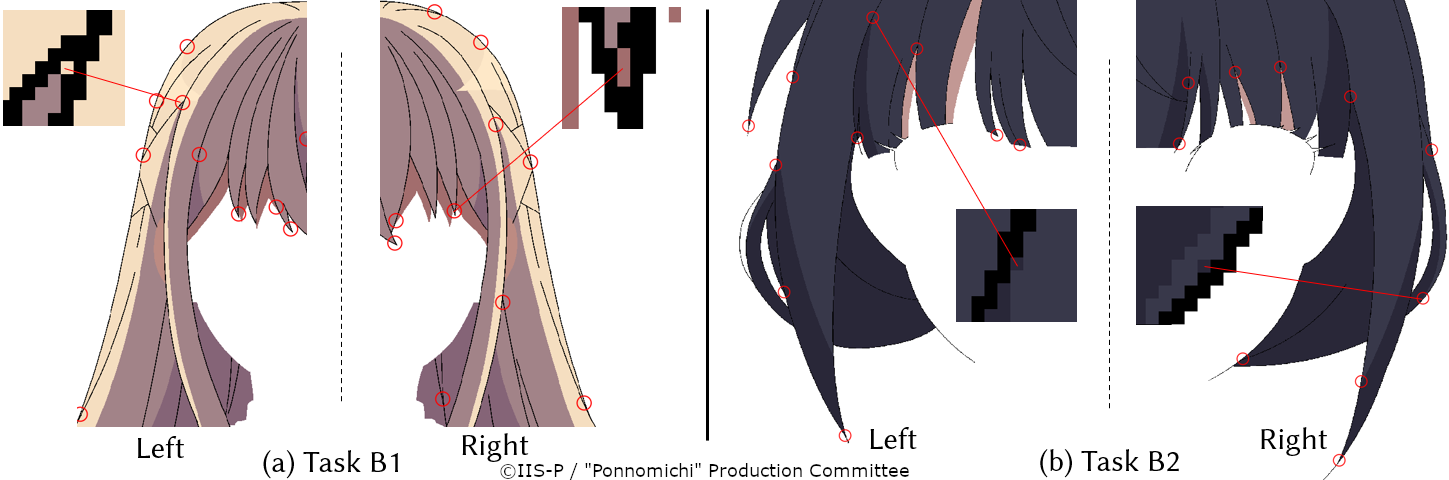}
  \caption[Task B image sets]{
Overview of Task B image sets ((a) B1 and (b) B2), with intentionally controlled gap settings. Red circles indicate unpainted gaps, and the magnified view highlights a color prediction error output by \systemName.
  }
  \Description{A figure displaying the two sets of test images used in ``Task B'', labeled (a) and (b).
Both images show an anime character's hair, symmetrically split into ``Left'' and ``Right'' halves to facilitate a comparative study.
Red circles indicate the locations of $10$ intentionally placed unpainted gaps (these circles were invisible to participants).
(a) ``Task B1'' represents an ``easier task'' using light-colored hair with distinct contours. A magnified inset shows a ``prediction error'' where \systemName incorrectly filled a gap with a mismatch color instead of the surrounding context.
(b) ``Task B2'' represents a ``more difficult task'' using dark hair with low color contrast. Its inset also highlights a prediction error.
The figure illustrates how the experiment controlled the number and difficulty of gaps, including specific cases where the AI makes mistakes.}
  \label{fig:taskb_fig}
\end{figure*}

Informed by our formative study, the painting software incorporated several fundamental tools common in professional colorization workflows (\textit{Paint Bucket}, \textit{Color Picker}, \textit{Dot Pen}, and \textit{Eraser}). It also supported basic operations, such as panning and zooming, modeled after CSP to closely simulate production environments. Note that CSP does not support the development of add-ons, which necessitated the implementation of our painting software. To compare \systemName with conventional gap-filling methods, the system featured two experimental conditions. The \textit{Baseline} (\cref{fig:overview_ui}(a)) condition included three conventional tools, namely \textit{Enclose and Fill}, \textit{Leftover Pen}, and \textit{Black Light Method}, which were found to be commonly used in current anime production. In contrast, the \textit{Ours} (\cref{fig:overview_ui}(b)) condition introduced the proposed tool, \systemName ~\footnote{Gap size threshold and highlight color setting were fixed during the study.}. To simulate actual workflow while maintaining simplicity, the software provided preset layers: \textit{Line Art} (black outlines), \textit{Guides} (red/blue guidelines), \textit{Coloring} (partially filled or empty), and a white background, along with a \textit{Reference Image} (model sheets). Participants could only manipulate the \textit{Coloring} layer, ensuring focus on colorization.

Specifically, we prepared three types of tasks for the user study (details are provided in \cref{sec:evaluation:method:task}):
\begin{itemize}
  \item Task A: Color a line art from scratch (TL: 150 seconds, 1 set)
  \item Task B: Detect and fill unpainted gaps in an almost fully colored image (TL: 90 seconds, 2 sets)
  \item Task C: Observe and evaluate results after applying an automated color prediction (TL: 30 seconds, 4 sets)
\end{itemize}
TL (Time Limit) denotes the maximum allowed duration, designed to simulate high-pressure production deadlines. Operations on the canvas were disabled until participants pressed \textit{Start}, after which they could begin the task. They were instructed to press \textit{Done} to stop the timer once they considered the task complete.

For the image sets, a multiple-response question in the formative survey S1 (\textit{``Where do unpainted gaps often occur?''}) showed that 95\% of respondents identified \textit{``hair tips''} as a common location. Accordingly, we selected images from professional anime data that included such cases\footnote{These images were not included in the model's training process.}.
The tasks were arranged to fit within the 60-minute session, in the order shown in \cref{fig:userstudy_flow}, and explanations were provided via screen sharing. The images for the example tasks are presented in~\cref{fig:teaser}. The overall task sequence was fixed (\textit{Task A} $\rightarrow$ \textit{Task B} $\rightarrow$ \textit{Task C}). Within \textit{Task B} and \textit{Task C}, the order of subtasks, the presentation order of target images (left vs.\ right), and the order of UI conditions (\textit{Baseline} vs.\ \systemName) were counterbalanced to mitigate learning effects.

\subsubsection{Task Details}
\label{sec:evaluation:method:task}

\paragraph{Task A}
The objective was to conduct a comparative experiment under an unbiased condition to evaluate \systemName and to observe how professional colorists use this tool within a production-like colorization setting. To mitigate learning effects while ensuring comparable levels of visual complexity, we selected images of anime characters' hair with near symmetry and divided them into left and right halves. As a result, we obtained two target images with seemingly equivalent complexity, as displayed in~\cref{fig:userstudy_ui_taska}. Unlike those in Task B, the images for Task A were used directly from the dataset, taking into account that both the locations of unpainted gaps and the timing of handling them may vary across individuals.

\paragraph{Task B}

The objective was to conduct a comparative experiment to evaluate \systemName in a context focused on gap detection and filling. Accordingly, participants were situated in a setting similar to the final quality check stage. Task B followed the same principle as Task A, in which images of characters' hair with near symmetry from professional data were used and split into halves. To avoid clear advantage for either UI condition, both the number and placement of unpainted gaps were intentionally controlled. Specifically, images were generated by randomly selecting 10 spatially dispersed enclosed regions as gaps, ensuring the inclusion of at least one clear color prediction error by \systemName. \cref{fig:taskb_fig} shows the images used in Task B, where red circles mark the unpainted gaps and the magnified view shows an prediction error (these indicators were not visible during the user study). Two sets of images were prepared: an easier task with clearly distinct colors and relatively smaller image size for Task B1 (\cref{fig:taskb_fig}a), and a more difficult task with similar colors and larger image size for Task B2 (\cref{fig:taskb_fig}b).

\paragraph{Task C}

\begin{figure*}[t]
  \centering
  \includegraphics[width=\linewidth]{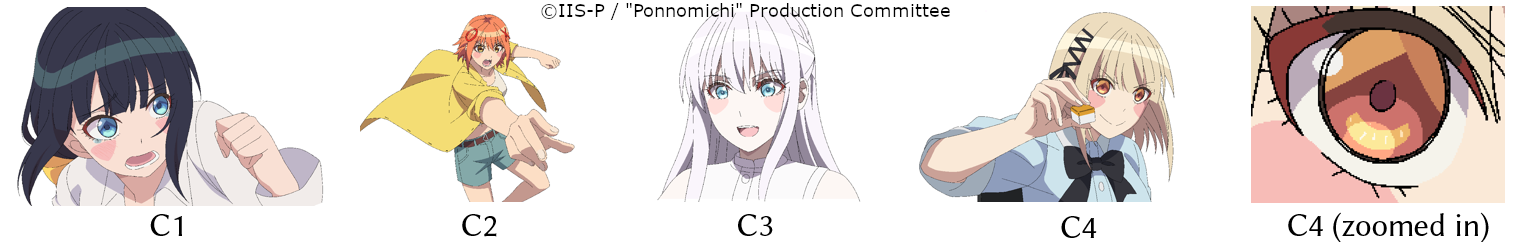}
  \caption[Task C image sets]{
Task C image sets (C1--C4). All enclosed regions below the threshold were filled with predicted colors; some mispredictions appear, for example, around the eyes.
}
\Description{The figure displays the set of four anime illustrations labeled ``C1'' through ``C4'' used in Task C, alongside a magnified detail view.
The illustrations depict female anime characters, full-size frames that contained complete character parts including
hair tips, with varying designs: ``C1'' has dark hair, ``C2'' has orange hair in an action pose, ``C3'' has silver hair, and ``C4'' has blonde hair.
A ``zoomed-in'' inset on the right highlights the eye region of ``C4''. It reveals the result of the system automatically filling small gaps, where some ``mispredictions'' are visible as incorrect colors in the intricate areas around the eye lines, demonstrating the challenges of processing complex real-world anime images.}
  \label{fig:taskc_fig}
\end{figure*}

The color prediction method was subjectively evaluated from the perspective of production. This was based on our hypothesis that, unlike color specifications provided by the model sheets, there is no single correct color for filling gaps, and any color may be acceptable as long as it appears appropriate.
Participants were instructed to inspect an image where gaps had been automatically filled by \systemName, assuming a quality check scenario before submission to the next production stage. They were given $30$ seconds to freely pan and zoom, after which they rated their impressions on a 7-point scale question CQ (\cref{sec:eval:sub}).
\cref{fig:taskc_fig} shows the images used for this task. These images were created by selecting full-size frames that contained complete character parts, including hair tips, identifying potential unpainted gaps as enclosed regions smaller than $3$ pixels, treating these as unpainted, and then applying our color prediction method in bulk to minimize arbitrariness.

\paragraph{Questionnaire and Interview for Combined Qualitative Feedback}
For the intermediate questionnaire via Google Forms, participants provided subjective evaluations of usability based on Tasks A and B. This questionnaire included several 7-point Likert-scale items~\cite{likert1932technique}, focusing on UI comparisons and the perceived usability of individual features of \systemName. The specific contents of questions (LQ1--LQ12) are presented in~\cref{sec:eval:eval}, where responses of $7$ are referred to as \textit{very positive}, and responses of $5$ or above are considered \textit{positive}. Participants were also required to submit the auto-saved coloring layer from each task, and asked to respond to four open-ended questions with OQ4 being optional:
\begin{itemize}
\item OQ1: Please provide any comments comparing the usability of the two UIs for handling unpainted gaps. %
\item OQ2: Regarding the overall usability of our UI (\systemName), please describe what you liked and what you think could be improved. %
\item OQ3: Were there any specific features of our UI (\systemName) that you found especially helpful, and why? %
\item OQ4: Please feel free to provide any additional comments regarding your experience with using our UI (\systemName).
\end{itemize}
At the end of the session, a semi-structured interview~\cite{lazar2017research} was conducted to further explore participants' impressions. This interview was guided by how they had used each tool in Tasks A and B, their questionnaire responses, and their evaluations from Task C, allowing us to obtain deeper insights. Feedback from the open-ended questionnaire responses and the interviews was analyzed together as combined qualitative data.

\subsection{Evaluation of \systemName}
\label{sec:eval:eval}

Following the ISO 9241-11~\cite{bevan2015iso} definition of usability, \systemName was evaluated based on three aspects: (1) \textit{task performance}, using objective measures such as task completion time (efficiency) and the number of overlooked unpainted gaps (effectiveness); (2) \textit{perceived usability}, using subjective measures (mainly satisfaction) such as questionnaire and interview responses; and (3) \textit{feature-level evaluation} conducted for each interaction (the color prediction method was separately evaluated in~\cref{sec:evaluation:prediction}).

\subsubsection{Task Performance}

\begin{figure*}[t]
  \centering
  \includegraphics[width=0.75\linewidth]{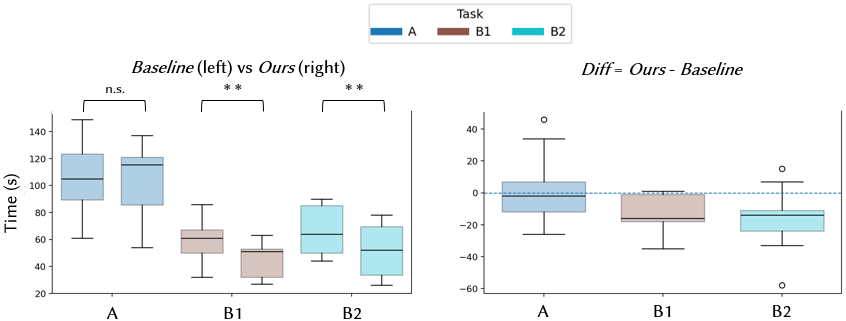}
    \caption[Task performance results]{
(a) Task completion times for each task (A, B1, and B2), and (b) their differences, summarized in boxplots.
    }
\Description{A pair of boxplot charts analyzing task completion times.
The left chart titled ``Baseline (left) vs Ours (right)'' compares the absolute time in seconds for three tasks: A, B1, and B2.
In Task A, the boxplots for Baseline and Ours are at similar heights, marked with ``n.s.'' (not significant).
In Tasks B1 and B2, the boxplots for ``Ours'' are positioned lower than ``Baseline'', indicating faster performance. These are marked with double asterisks (``**'') to show statistical significance.
The right chart titled ``Diff = Ours - Baseline'' shows the distribution of time differences.
For Task A, the boxplot is centered on the zero line.
For Tasks B1 and B2, the boxplots are primarily located below the zero line, demonstrating a consistent reduction in time when using \systemName.}
  \label{fig:taskab_performance}
\end{figure*}

To compare the efficiencies of the \textit{Baseline} UI and \systemName, we analyzed task completion times (recorded in seconds) using a within-subject design. Outliers were identified and excluded using Tukey's fences ($k = 1.5$)~\cite{tukey1977exploratory}. We primarily employed a one-sided Wilcoxon signed-rank test ($\alpha = .01$)~\cite{woolson2007wilcoxon} to evaluate the hypothesis that \systemName is faster (i.e., $H_1$: $\mu_{\text{Diff}} < 0$). Supplementary analysis using paired $t$-tests and visualizations of individual paired differences are detailed in Appendix B. In addition to efficiency, the effectiveness of these UIs was compared by the number of unpainted gaps remaining after task completion.

\paragraph{Efficiency}
As shown in~\cref{fig:taskab_performance}, for Task A ($n = 12$, P3 excluded), no significant difference was found between the \textit{Baseline} ($M = 106.50, SD = 26.65$) and \systemName ($M = 104.17, SD = 25.15$) with $W = 28, p = .212$. However, significant improvements were observed in Task B. In Task B1 ($n = 13$, no outliers), \systemName ($M = 45.15, SD = 12.28$) was faster than the \textit{Baseline} ($M = 57.69, SD = 15.59$) with $W = 6, p = .003$. Similarly, in Task B2 ($n = 11$, P6 and P7 excluded), \systemName ($M = 51.91, SD = 19.66$) outperformed the \textit{Baseline} ($M = 66.27, SD = 18.46$) with $W = 3, p = .002$.

\paragraph{Effectiveness} \systemName resulted in zero unpainted gaps across all participants and tasks. In contrast, in the \textit{Baseline} condition, unpainted gaps remained for specific participants: P4 left one gap in Task A; P6 and P10 left one gap each in Task B1; and in Task B2, P6 left five gaps, P11 and P14 left two gaps each, and P12 left one gap.

\paragraph{Summary} In conclusion, at a significance level of $\alpha = .01$ (one-sided), \systemName was significantly faster than the \textit{baseline} UI in Tasks Bs. Regarding remaining gaps, \systemName consistently outperformed the \textit{baseline} tool across all tasks. These results indicated that \systemName is more efficient and effective than conventional practices, particularly when detecting and filling gaps is critical, whereas its performance in coloring from scratch may vary across individuals. In fact, P8 remarked that \textit{``Auto-detecting the fill gap definitely saves a lot more time than using the black light, which is good for final checking. But when it comes to initial painting, the baseline UI is a bit easier since I can just use the Leftover Pen to cover up gaps while painting. Overall, both have their advantages, and \systemName does save a lot of work time during the final check.''}, reflecting his workflow in Task A, where he used the \textit{Leftover Pen} to color multiple regions at once, leaving only a few unpainted gaps. Furthermore, P2, P7, P9, and P14 suggested a combination of both tools. A deeper discussion of the potential for integrating both tools is provided in~\cref{sec:discussion:conbination}.

\begin{figure*}[t]
  \centering
  \includegraphics[width=0.875\linewidth]{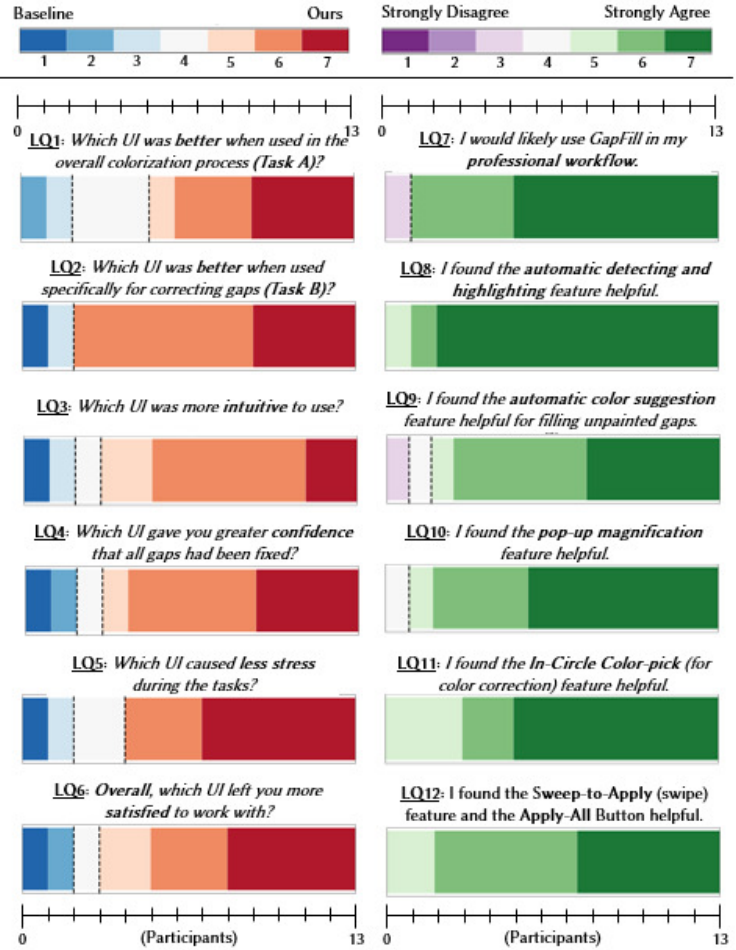}
    \caption[Likert results]{
Summary of responses to the 7-point Likert-scale items. LQ1–7 evaluate perceived usability, whereas LQ8–12 evaluate individual features.
    }
\Description{A summary of questionnaire responses from 13 participants presented as stacked bar charts.
The charts are divided into two categories.
The left column (LQ1--LQ6) evaluates perceived usability on a 7-point scale from ``Baseline'' (blue) to ``Ours'' (red). The results show a dominant shift towards the red side, indicating that participants rated the proposed method significantly higher in terms of overall process, gap correction, intuitiveness, confidence, stress reduction, and satisfaction. (The specific questions are as follows. LQ1: Which UI was better when used in
the overall colorization (Task A)?
; LQ2: Which UI was better when used specifically for correcting gaps (Task B)?; LQ3: Which UI was more intuitive to use?; LQ4: Which UI gave you greater confidence that all gaps had been fixed?; LQ5: Which UI caused less stress during the tasks?; LQ6: Overall, which UI left you more satisfied to work with?)
The right column (LQ7--LQ12) evaluates individual features on a scale from ``Strongly Disagree'' (purple) to ``Strongly Agree'' (green). These bars are almost entirely filled with shades of green, demonstrating that participants strongly agreed that the automatic detection, highlighting, color suggestions, pop-up magnification, and interaction techniques were helpful and suitable for their professional workflow. (The specific questions are as follows. LQ7: I would likely use GapFill in my professional workflow.; LQ8: I found the automatic detecting and highlighting feature helpful.; LQ9: I found the automatic color suggestion feature helpful for filling unpainted gaps.; LQ10: I found the pop-up magnification feature helpful.; LQ11: I found the In-Circle Color-pick (for color correction) feature helpful.; LQ12: I found the Sweep-to-Apply (swipe) feature and the Apply-All Button helpful.)}
  \label{fig:likert_usability}
\end{figure*}

\subsubsection{Perceived Usability}

To assess subjective usability, we designed the 7-point Likert-scale questions to compare UIs.
\cref{fig:likert_usability} summarizes responses to LQ1–-LQ7. Overall, both subjective ratings and participant feedback indicated strong support for \systemName, particularly in detecting gaps. Regarding usability in each task, eight participants rated \systemName more positively than \textit{Baseline} in LQ1, and eleven did so in LQ2. This aligns with the performance-based evaluation and the generally favorable feedback from participants (ten positive in LQ6; \textit{``It serves what it is supposed to do.''} (P5)). However, P3, who consistently preferred the \textit{Baseline}, noted that \textit{``I didn’t trust AI so much actually […] maybe this is my first time using the AI tool''}, indicating a preference for manual tools.

Regarding more fine-grained usability aspects, ten participants responded positively in LQ3 on intuitiveness (\textit{``Overall process is very straightforward and simple.''} (P2)), although P3, P9, and P14 also noted the need to adapt to a new tool (\textit{``I think I will use it because it is easier, but it might take a bit of time to adapt since I am already used to zooming in.''} (P14)). Similarly, ten participants were positive in LQ4 regarding confidence in filling gaps (\textit{``With the Baseline UI, even though you think you filled everything, you cannot always see it properly. With Ours, the AI basically detects the ones that have not been filled.''} (P10)). This was consistent with observations in Task B2: although P6 and P10 had finished filling all the gaps with \textit{Baseline}, they could not immediately recognize whether any gaps remained and hit the time limit. However, P9 expressed stronger trust in conventional methods, remarking that \textit{``\systemName needs to have the Black Light mode to check if the AI properly detected and filled all the holes for the final check.''}, suggesting a practical consideration for new AI-powered tools. Regarding reduced stress, nine participants were positive in LQ5 (\textit{``\systemName really helps […] because the Baseline UI relies on our own perception to catch the gaps''} (P5)).

In terms of practical adoption (LQ7), all participants except P3 responded positively, with eight being strongly positive. P11 emphasized its benefit under time-pressured situations: \textit{``If you are late on a deadline, \systemName can save a lot of time identifying mistakes.''} Altogether, \systemName was rated more favorably than \textit{Baseline} in perceived usability, particularly for detecting gaps. A more detailed discussion on integrating the advantages of both tools for practical deployment is provided in~\cref{sec:discussion:conbination}.

\subsubsection{Feature-Level Evaluation}

To evaluate the usability of each individual feature, we designed a set of 7-point Likert-scale questions.
A summary of responses to LQ8--LQ12 is presented in~\cref{fig:likert_usability}, while a detailed analysis of each feature is provided in the following paragraphs. The contribution of color suggestion accuracy to overall usability is discussed separately in~\cref{sec:discussion:accuracy}.

\paragraph{Unpainted Gap Detector with Circular Highlights}
Based on LQ8, all participants except P3 and P9 were very positive and all participants were positive toward this feature. According to the combined qualitative feedback, participants consistently appreciated the automatic circle-based highlighting of gaps, which enabled faster detection and allowed them to readily verify whether any unpainted areas remained (\textit{``usually you need to always zoom in and out and sometimes it feels like not confident that we have fix all, the circle is really helpful''} (P14)). This was regarded as a clear improvement over the conventional method of repeatedly toggling the \textit{Black Light} while zooming and panning across the canvas to ensure no gaps were missed (\textit{``We tend to rely on multiple times checking […] Having a tool to help with this process lessen the burden […] With very detailed characters […] this one actually is going to be very helpful in the future''} (P5)). While the overall impressions were satisfactory, P8 also suggested improvements regarding clustered cases: \textit{``need to improve on part where there are multiple unfilled gaps on same spot […] sometimes quite confusing which circle is for which gaps.''}

\paragraph{Automatic Color Suggestion}
Based on LQ9, six participants were very positive, and all participants except P4 and P12 were positive toward this feature. P2, P6, and P8 valued the automatic color-filling function, which reduces the effort of manual color selection (\textit{``I like how the Ours UI immediately applied the colors that I wanted when fixing the gaps, making the work less time-consuming''} (P6)). In contrast, P10 pointed out that \textit{``the AI color suggestion should be a little bit more noticeable in terms of where the color was sourced,''} suggesting the possibility of reflecting the discrete nature of the problem setting in the UI design.

\paragraph{Hover-Activated Pop-up magnification}
Based on LQ10, seven participants were very positive and all participants except P12 were positive toward this feature. P6–P10 and P14 valued the concept of displaying a magnified view on hover, as it reduced the need for constant zooming (\textit{``the pop-up map is helpful to see the coloured gaps without always zooming in.''} (P14)). However, participants also pointed out room for improvement in the magnification settings. For example, P1 noted, \textit{``When the color shade difference doesn't look obvious with the zoomed UI […] still need to look closely to make sure what color it fills in.''} Likewise, P4 remarked, \textit{``the size is too small. Usually the red marker blends with the color inside the square. So sometimes you kind of miss see the color and then that's why even if it predicted correctly, I was trying to switch it around,''} both providing suggestions for future development.

\paragraph{In-Circle Color-pick}
Based on LQ11, eight participants were very positive and all participants were positive toward this feature. P5--P8, P10, and P11 valued the ability to manually change the AI-suggested color via an intuitive interaction (\textit{``I like the pop-up one where we can just drag to easily pick nearby colour and fill the gap.''} (P7)). While P8 appreciated the dotted line for clarifying the substitution target, participants also suggested possible improvements. For instance, P4 commented that \textit{``in-circle color-picker could just be a hover-to-select rather than a drag-to-select one,''} and P11 remarked that \textit{``for the color picker I prefer a color code show like the RGB code […] Because when the color is very close to each other it is so confusing,''} both pointing to directions for future development.

\paragraph{Out-Circle Sweep-to-Apply and Apply-All Button}
Based on LQ12, six participants were very positive and all participants were positive toward these features, appreciating the ability to apply colors at once instead of one by one. However, the strategy for applying predicted colors during the task varied. For example, P1, P2, P5, P8, P9, and P12 primarily used the magnified view to check and correct colors one by one, and then applied them all at once at the end with the \textit{Apply-All button} (\textit{``I'm the guy to like just press one button of one motion''} (P1)). In contrast, P3, P10, and P11 tended to apply them in groups of clustered or neighboring highlights using the \textit{Sweep-to-Apply} feature (\textit{``I very like Sweep-to-apply, it is very convenient and fast''} (P3)). The remaining participants flexibly combined both approaches depending on the task, suggesting that the most appropriate application method may vary according to the number and distribution of unpainted gaps. However, P7 mentioned that \textit{``What could be improve may be the Sweep-to-Apply, I think like I missed to pick the colour sometimes''}, which reflects their dissatisfaction with applying clustered highlights and subsequently having to correct the color manually using the \textit{Paint Bucket} tool.

\subsection{Evaluation of Our Color Prediction Method}
\label{sec:evaluation:prediction}

\subsubsection{Subjective Evaluation}
\label{sec:eval:sub}

\begin{figure}[t]
  \centering
  \includegraphics[width=0.85\linewidth]{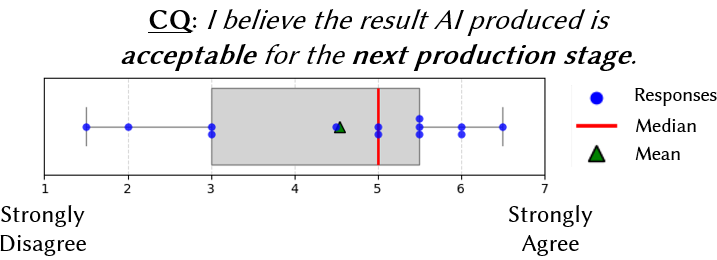}
    \caption[CQ result]{Summary of responses to CQ (production acceptability), based on the median across four images (C1–-C4).
    }
\Description{A box plot with overlaid individual data points (blue dots) summarizing participant responses to the statement: ``I believe the result AI produced is acceptable for the next production stage.''
The x-axis represents a 7-point Likert scale ranging from $1$ (``Strongly Disagree'') to $7$ (``Strongly Agree'').
Key statistics are shown visually: the median (red line) is positioned at $5$, indicating a generally positive acceptance. The mean (green triangle) is slightly lower, at approximately $4.5$.
The interquartile range (gray box) spans from $3$ to $5.5$, showing a spread of opinions. While some individual responses fall below $2$ (disagreement), the central tendency suggests that the AI's output was considered acceptable by many participants.}
  \label{fig:taskc_res}
\end{figure}

To avoid pseudoreplication~\cite{hurlbert1984pseudoreplication}, we used the median of responses across four images (C1--C4, shown in~\cref{fig:taskc_fig}) as the representative score for each participant. The aggregated results are shown in~\cref{fig:taskc_res}.
For production acceptability (CQ), the responses were $M = 4.5, SD = 1.6, Md. = 5$. The results indicate that our method received positive evaluations subjectively from a production perspective. However, the fact that direct application of predictions is impermissible underscores the strict standards of professional colorization. Qualitative feedback further supported these findings. P1, P2, P6, and P8--P12 appreciated the overall high accuracy, though they consistently noted that corrections were needed around the pupil (\cref{fig:taskc_fig}), which is composed of small regions with diverse colors (\textit{``I think it’s accurate when the color is like in a large area. But like for the eyes it will have some problem''} (P1)). However, P2, P6, P9, and P11 also emphasized that these corrections were minor and consistently localized, making them easy to address (\textit{``the part that requires fixing is so little and always the same part of the picture''} (P11)).

In considering whether anime viewers would even notice such minor inconsistencies, P6 reflected: \textit{``For me, I don't think they will notice because the images after the production will end up being blurry […] In my opinion still very important to make things very neatly for our satisfaction and also for the viewers but if we are in a last minute rush then I think we can just give it to the next. Depends on the schedule.''} This perspective illustrates that while small inaccuracies may not significantly impact the viewing experience, there remains a professional expectation for precision, balanced against practical production constraints. Taken together, these insights reinforce the acceptability of our method in production workflows. A more detailed discussion on how prediction accuracy for small gaps contributes to the usability of \systemName is provided in~\cref{sec:discussion:accuracy}.

\subsubsection{Objective Evaluation}

\begin{figure}[t]
  \centering
  \includegraphics[width=\linewidth]{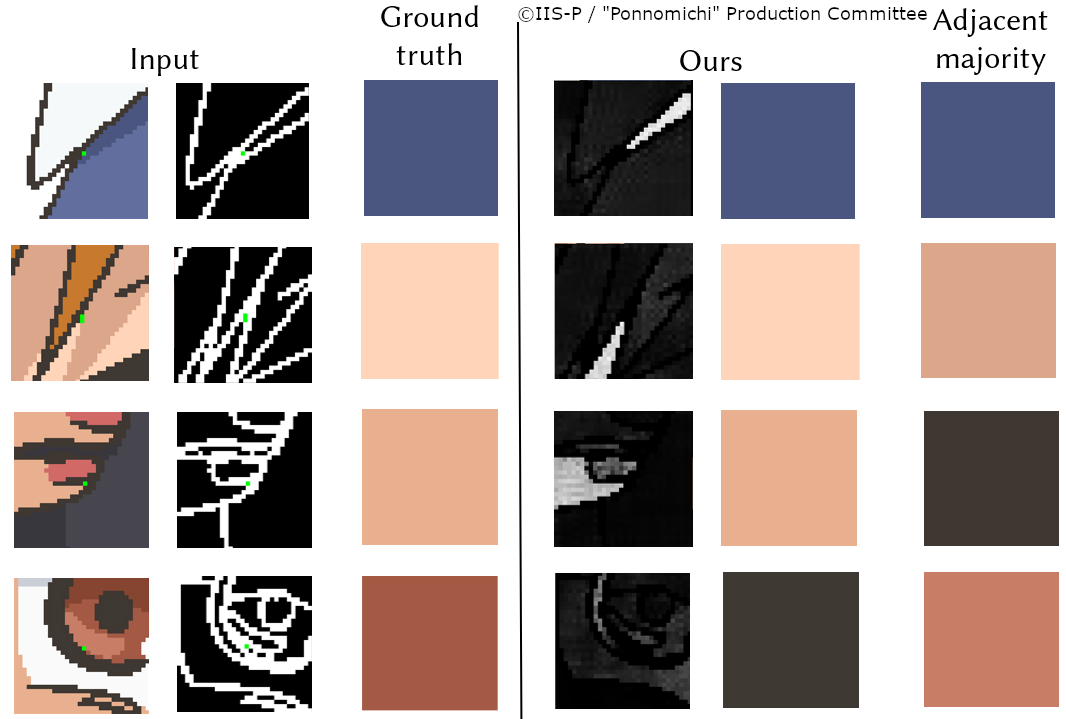}
    \caption[Color prediction evaluation]{Example comparison of color predictions for gaps (highlighted in green) between our method (with likelihood map) and a rule-based adjacent majority approach.
    }
\Description{A visual comparison of color prediction results across four different examples (rows), comparing ``Input'', ``Ground truth'', ``Ours'', and ``Adjacent majority''.
The columns display the input patch with a target gap (green dot), the correct color, the proposed method's likelihood map and result, and the baseline method's result.
Top rows (1--2): In simpler boundary cases like hair tips, both methods generally succeed in row 1, but sometimes the ``Adjacent majority'' incorrectly picks a different color.
Bottom rows (3--4): In complex regions, the methods diverge significantly.
Row 3 shows a gap near a mouth line; ``Adjacent majority'' incorrectly picks the dark hair color dominating the neighborhood, while ``Ours'' correctly predicts the pinkish skin tone.
Row 4 shows a gap inside a detailed eye line; ``Adjacent majority'' incorrectly picks the surrounding skin color, while ``Ours'' also incorrectly identifies the dark brown of the eye line.
This demonstrates the proposed method's ability to handle intricate structures where simple adjacency rules fail.}
  \label{fig:eval_prediction}
\end{figure}

We used a professional anime title that was not included in the training set and selected a single episode from it for objective evaluation. Unpainted gaps were defined as regions containing fewer than $10$ pixels (same as the training process), and accuracy was measured as the proportion of cases in which the predicted color exactly matched the ground truth. Since most existing automatic anime colorization methods require reference images, we compared our approach against a naive baseline: a rule-based greedy method that selects the majority color among the eight-connected neighboring pixels of the target region (\textit{Naive}).

Our method achieved an accuracy of $81.68\%$ ($83{,}345/102{,}041$), whereas the \textit{Naive} baseline achieved $37.02\%$ ($37{,}776/102{,}041$), as illustrated in~\cref{fig:eval_prediction}. These results demonstrate the effectiveness of our approach for color suggestion in unpainted gaps. Moreover, the average inference time per patch was $74 \mathrm{ms}$ on an NVIDIA RTX 6000 Ada Generation, indicating that the prediction speed is sufficient to support real-time interaction.

\section{Discussion and Future Work}
\label{sec:discussion}

\subsection{Contribution of Color Prediction Accuracy to Production Tool Usability}
\label{sec:discussion:accuracy}

Results showed that \systemName has high usability for gap-filling; participants consistently valued features such as clear gap visualization and user control over AI suggestions. However, as P5 remarked:
\textit{``for the \systemName […] it's already completed because it does what it supposed to do, and for the AI one for predicting the colors I would say I think it's pretty good but some of it still need to be minor fix by the painter,''} suggesting that the accuracy of color prediction should be discussed independently in assessing usability. This distinction aligns with the recommendation by ~\citet{remy2020evaluating} to clearly define the goals and factors of evaluation for CSTs.

\subsubsection{Accuracy as a Conditional Factor}
Participants expressed differing views on how color prediction accuracy affects the usability of \systemName. Some considered errors acceptable if they could be easily corrected. For example, P9 stated, \textit{``There is some wrong colors […] but it can be easily fixed using \systemName, just pick color then done.''}, and P5 found the ability to manually modify AI outputs satisfying. This aligns with broader Human-AI collaboration literature indicating that granting decision control to users mitigates the impact of AI errors~\cite{singh2025systematic}, thereby fostering acceptance of imperfect systems~\cite{kocielnik2019will} and user trust~\cite{westphal2023decision}.

In contrast, others emphasized that even small mistakes undermined trust: P3 noted, \textit{``when AI makes a mistake […] you can't trust it anymore. I am scared there's a wrong part,''} while P10 stressed the need for greater consistency in its accuracy. This reflects \textit{algorithm aversion}~\cite{dietvorst2015algorithm}, where observing algorithmic errors causes a sharper confidence decline than human errors. Such reactions align with automation trust research suggesting that even a single visible mistake, particularly in easy tasks, can substantially reduce future trust~\cite{madhavan2006automation, centeio2023exploring}, stressing the significance of observed accuracy~\cite{yin2019understanding}.

A more conditional stance was also proposed. P2 noted that results were generally reliable but still required double-checking around sensitive regions. Similarly, P6 and P11 pointed out that whether predictions could be accepted without correction often depends on production demands. This reality reflects the environmental \textit{Press} in \citet{rhodes1961analysis}'s 4P model, where external constraints influence decision-making behaviors. Specifically, tight deadlines force practitioners to navigate a strategic accuracy-time tradeoff~\cite{swaroop2024accuracy}, compromising the quality of visual inspection~\cite{rieger2022human} while leaving insufficient final decision time for manual verification~\cite{cao2023time}.

In summary, prediction accuracy may play a conditional role in usability: while sufficient accuracy makes the tool practically valuable, its impact could be less than manual control over AI's output and depends on both task sensitivity and production context. Minor errors are often tolerable if easily corrected, but inaccuracies in critical areas can erode trust. In other words, prediction accuracy alone is unlikely to determine the tool’s usability; rather, it is the combination with other features and contextual factors that matters.

\subsubsection{Redefining ``Accuracy'' in Practice}

\begin{figure}[t]
  \centering
  \includegraphics[width=\linewidth]{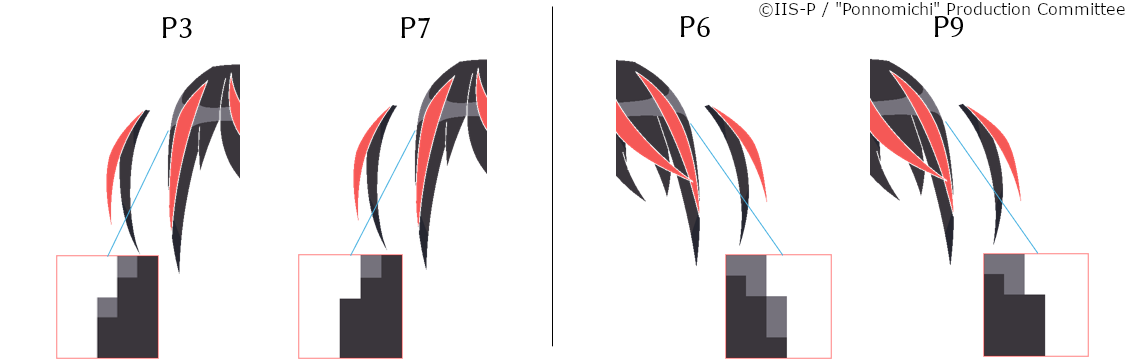}
    \caption[Gap color diffs]{Coloring results from Task A. As highlighted in the magnified view, participants occasionally used different colors when filling gaps.
    }
\Description{A comparison of coloring results from four participants (P3, P7, P6, and P9) showing inconsistencies in manual color selection for gaps during ``Task A''.
The figure presents two pairs of identical hair segments.
The left pair compares P3 and P7. A red square ``magnified view'' zooms in on a specific gap at the tip of a hair strand. P3 filled these pixels with a lighter gray shadow color, whereas P7 filled the exact same pixels with a dark black color.
The right pair compares P6 and P9, showing a similar discrepancy where P6 used a lighter shade than P9 for the same gap location.
This illustrates that even for the same image, different professionals may make different judgments on which neighboring color should be used to fill a gap.}
  \label{fig:gap_diff}
\end{figure}

\cref{fig:gap_diff} shows that small gaps were occasionally filled with different neighboring colors across participants in Task A, yet the outcomes were visually indistinguishable. This raises the question of whether pixel-level accuracy is an appropriate metric for evaluation, and whether a definitive ground truth can be assumed in such cases, unlike standardized color specifications from model sheets. Indeed, some participants (P6, I1, and I4) believed that most anime viewers would not notice such subtle inconsistencies after post-production, and several prior studies have pointed out that straightforward error metrics are sometimes insufficient for assessing human perceptual quality~\cite{wang2009mean, zhang2018unreasonable}. These observations suggest that minor errors in inconspicuous areas may have little impact on either production quality or viewer perception. 

Consequently, the contribution of prediction accuracy must be reconsidered in terms of where and when errors occur. If a gap is tiny and visually blended, any neighboring color may suffice, whereas inaccuracies at sharp boundaries or in semantically important regions (e.g., the pupil) can seriously harm perceived quality. This perspective shifts the focus from absolute accuracy metrics to situational accuracy criteria that better capture real-world demands in production. This also aligns with the concept of recent co-creative AI frameworks, which argues that beyond algorithmic accuracy effective collaboration depends on high-quality, contextually appropriate interaction~\cite{rezwana2023designing, rezwana2021cofi}.

\subsubsection{Future Directions: Technical Improvements and Reconsidering Accuracy Definitions}
First, from a technical perspective, improving accuracy in detailed regions, particularly the pupils, remains an important challenge. Potential solutions include leveraging reference images and neighboring frame information to enhance performance. Moreover, enabling robust prediction even when adjacent areas are sparsely colored would improve flexibility, surpassing current assumptions that require local context to be mostly filled. Supporting dynamic updates that reflect user-selected fill colors is also a key objective for enhancing usability. 

Second, reconsidering the problem setting is essential to clarify when prediction accuracy for small gaps truly matters. While improving pixel-level accuracy for tiny regions is known to be difficult, it may be less critical if the visual result is perceptually identical to the ground truth. However, mistakes in critical regions, such as inside the eyes or at strong color boundaries, significantly degrade quality. Therefore, future investigations should examine conditions such as gap size, position, and surrounding contrast to determine viewer tolerance for approximate results. Defining these criteria will guide the design and evaluation of more effective AI-assisted anime colorization methods to meet real-world requirements.

\subsection{Trust in AI-Powered Assistance for Colorists: Integration with Existing Tools}
\label{sec:discussion:conbination}
Our study demonstrated that \systemName offers advantages for gap-filling tasks. The high appraisal for freely toggling \systemName and visualizing gaps via highlighting or magnified views aligns with the strategy of ``view-shifts between component and composition'' for creativity experts~\cite{frich2019strategies}. However, it also raises important questions about how professionals situate such new AI-powered tools within established practices and the extent to which they trust them.

\subsubsection{Limits of Trust in New AI-Powered Assistance}
Despite the benefits of \systemName, some participants hesitated to rely solely on AI output. P3 double-checked their work even after all highlights disappeared due to concerns about AI reliability, P7 often manually selected the same color suggested by the AI, and P9 required the \textit{Black Light} for a final check after using \systemName. These behaviors mirror friction in human collaborations, where establishing ``trust in skills'' precedes relinquishing control to a support actor~\cite{chung2022artist}, and likely reflect both limited professional familiarity with such tools and the high accountability required of colorists, consistent with observations that creative workers adopt emerging AI cautiously while balancing benefits against uncertainties~\cite{vimpari2023adapt}. This resonates with process views of human-AI collaboration, where trust must be actively managed over time rather than assumed after adoption~\cite{mcgrath2025collaborative}.

Trust issues were also salient in visually sensitive regions, where minor mistakes can affect a character’s impression. As P4 noted, \textit{``if you miscolor the eye […] the shape will not be correct […] so usually we make sure the eye actually has the proper shape''}, and concluded, \textit{``It can be the tool to exist but it cannot be the final product.''} This echoes reports that AI outputs often lack nuance and require human expertise for final quality assurance~\cite{inie2023designing}. Likewise, \citet{grassini2025artificial} show that while AI can produce semantically diverse ideas, human creative output is perceived as higher quality and nuance. Overall, these remarks underscore that while AI assistance can be useful, final accountability remains with human hands.

\subsubsection{Toward Integration with Existing Tools}
Crucially, several participants envisioned combining the two systems. P7 and P8 valued each system’s strengths depending on the task, concluding that combining both would be most useful. This aligns with~\citet{palani2024evolving}, who found that creative workers prefer AI that supports tasks without disrupting ownership of their familiar process. Likewise, P2 remarked, \textit{``Best to combine […] I’m already used to painting with the brush-fill method, so not having that kind of slows my progress a bit.''}, underscoring the transition cost between methods. In fact, AI-powered assistance can create friction when it conflicts with established routines~\cite{ogawa2025understanding} or increases the communication and interpretation burden associated with agentic AI outputs~\cite{gmeiner2023exploring}, highlighting the need to embed such systems into existing workflows and meet the role expectations.

These perspectives reflect a pragmatic stance: rather than replacing established practices, AI-powered tools may be most valuable when flexibly integrated into existing workflows, allowing professionals to benefit from the speed of automation while retaining trusted manual verification. Similarly, ~\citet{uusitalo2024clay} describe generative AI in design as ``clay to play with'', a medium for rapid experimentation that preserves human authorship and control. Complementing this, ~\citet{popova2023co} advocate embedding AI into toolchains to augment established practices, supported by ongoing quality assurance and user education. From a broader workflow perspective, our findings suggest that \systemName has the potential to seamlessly complement existing tools by leveraging their strengths.

\subsubsection{Future Directions: User Interface Improvements} 
\begin{figure}[t]
  \centering
  \includegraphics[width=0.75\linewidth]{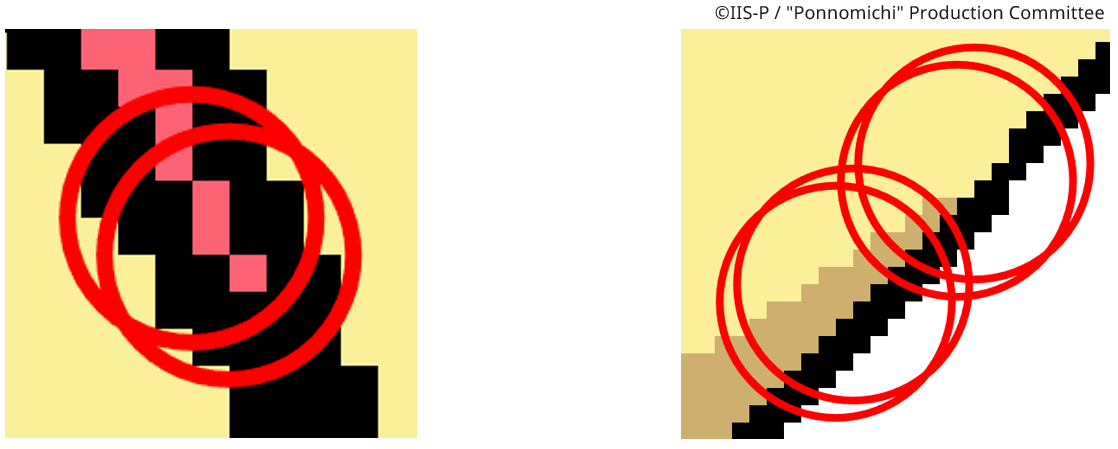}
    \caption[Gap clusters]{Dense gap clusters cause visual clutter, with overlapping circles obscuring gap–highlight correspondence.}
\Description{Two zoomed-in examples illustrating the issue of ``visual clutter'' caused by dense clusters of gaps.
In both images, multiple unpainted gaps are detected in close proximity along the edges of line art.
As a result, the red highlight circles generated for each gap overlap heavily with one another.
This overlapping creates a confused visual state where it is difficult for the user to distinguish individual gaps or determine which highlight corresponds to which specific pixel correction.}
  \label{fig:future_cluster}
\end{figure}

Some professionals suggested that our system would be more readily adopted in production workflows if it were integrated with established practices such as \textit{Black Light Method}, which are already familiar and trusted. Combining our method with other tools and leveraging their strengths at different stages of the process could further enhance usability. Thus, a tool designed to complement rather than replace existing methods may be crucial for practical adoption. In addition, usability can benefit from refinements to minor interface settings of \systemName, particularly those related to gap visualization. For example, clustered gaps (\cref{fig:future_cluster}) should be highlighted in a way that reduces overlap to improve the correspondence between gaps and highlights, and the temporary fill color should be made more distinguishable. Moreover, extending the system to handle unpainted gaps caused by anti-aliased line art (not only binary lines), would expand its applicability beyond anime production.

\section{Conclusion}
\label{sec:conclusion}

We addressed the underexplored challenge of small unpainted regions (``gaps'') in professional anime colorization workflows. Building on a formative study with industry practitioners, we developed \systemName, a specialized tool that integrates seamlessly into existing pipelines and reduces the effort required for gap detection, zooming, and color selection, supported by a deep learning method that leverages the flat-color characteristics of anime images and local cues. A user study with $13$ professional colorists demonstrated significant efficiency gains in gap-filling tasks compared to existing tools, while also showing that usability may be driven by the combination of automatic detection and clear visualization of gaps, as well as user control over AI suggestions rather than prediction accuracy alone. We further found that appropriate gap-filling colors can be contextually ambiguous. From a broader workflow perspective, our findings suggest that \systemName has the potential to integrate with existing tools by leveraging their respective strengths, while also providing insights into professionals' trust in new AI-powered assistance. Taken together, these contributions provide practical value for anime production and broader insights into the design of AI-assisted creativity support tools that respect established practices while aiming at their gradual adoption in production contexts.

\begin{acks}
This paper is based on results obtained from GENIAC (Generative AI Accelerator Challenge, a project to strengthen Japan’s generative AI development capabilities), a project implemented by the Ministry of Economy, Trade and Industry (METI) and the New Energy and Industrial Technology Development Organization (NEDO), Japan Grant Number  JPNP20017. This work was also supported by JST, CRONOS, Japan Grant Number JPMJCS25K1. Finally, we thank the professional colorists at OLM Asia SDN BHD for their participation and valuable feedback in our formative and user studies, and we also express our sincere gratitude to the members of \copyright IIS-P/Ponnomichi Production Committee for kindly providing the image data.
\end{acks}

\bibliographystyle{ACM-Reference-Format}
\bibliography{reference}

@inproceedings{kato2024griffith,
  title={Griffith: A Storyboarding Tool Designed with Japanese Animation Professionals},
  author={Kato, Jun and Hara, Kenta and Hirasawa, Nao},
  booktitle={Proceedings of the 2024 CHI Conference on Human Factors in Computing Systems},
  pages={1--14},
  year={2024}
}

@article{ichikohji2013influence,
  title={The Influence of Introducing IT into Production System A Case of Japanese Animation (Anime) Industry},
  author={Ichikohji, Takeyasu},
  journal={Annals of Business Administrative Science},
  volume={12},
  number={4},
  pages={181--197},
  year={2013},
  publisher={Global Business Research Center}
}

@book{otsuka_2022_animemade,
  author    = {Otsuka, Takashi and Hotta, Takayuki and Funayama, Yasuaki},
  title     = {Anime ga Dekiru made [The Making of Anime]},
  publisher = {Asukashinsha},
  year      = {2022},
  asin      = {B0BLYKKTC5},
  language  = {japanese}
}

@book{yokota_2019_psychology,
  editor    = {Yokota, Masao},
  title     = {Animēshon no Shinrigaku [The Psychology of Animation]},
  series    = {Nihon Shinri Gakkai Shinrigaku Sōsho [The Japanese Psychological Association Monograph Series on Psychology]},
  publisher = {Seishin Shobō},
  location  = {Tokyo},
  year      = {2019},
  isbn      = {978-4414311235},
  language  = {japanese}
}

@inproceedings{kato2025anime,
  title={Anime SIG: Researching Japanese Animation From Technical, Cultural, and Industrial Perspectives},
  author={Kato, Jun and Koyama, Yuki and Maejima, Akinobu and Mihara, Ryotaro and Seaborn, Katie},
  booktitle={Proceedings of the Extended Abstracts of the CHI Conference on Human Factors in Computing Systems},
  pages={1--3},
  year={2025}
}

@book{condry2013soul,
  title={The soul of anime: Collaborative creativity and Japan's media success story},
  author={Condry, Ian},
  year={2013},
  publisher={Duke University Press}
}

@incollection{todo2024practical,
  title={A Practical Style Transfer Pipeline for 3D Animation: Insights from Production R\&D},
  author={Todo, Hideki and Koyama, Yuki and Sakai, Kunihiro and Komiya, Akihiro and Kato, Jun},
  booktitle={SIGGRAPH Asia 2024 Technical Communications},
  pages={1--4},
  year={2024}
}

@inproceedings{wang2024apisr,
  title={Apisr: Anime production inspired real-world anime super-resolution},
  author={Wang, Boyang and Yang, Fengyu and Yu, Xihang and Zhang, Chao and Zhao, Hanbin},
  booktitle={Proceedings of the IEEE/CVF Conference on Computer Vision and Pattern Recognition},
  pages={25574--25584},
  year={2024}
}

@article{mihara_2020_coming,
  author  = {Mihara, Ryotaro},
  title   = {{A Coming of Age in the Anthropological Study of Anime?: Introductory Thoughts Envisioning the Business Anthropology of Japanese Animation}},
  journal = {Journal of Business Anthropology},
  year    = {2020},
  volume  = {9},
  number  = {1},
  pages   = {88--110}
}

@article{morisawa_2015_managing,
  author  = {Morisawa, Tomohiro},
  title   = {{Managing the unmanageable: Emotional labour and creative hierarchy in the Japanese animation industry}},
  journal = {Ethnography},
  year    = {2015},
  volume  = {16},
  number  = {2},
  pages   = {262--284}
}

@article{qu2006manga,
  title={Manga colorization},
  author={Qu, Yingge and Wong, Tien-Tsin and Heng, Pheng-Ann},
  journal={ACM Transactions on Graphics (ToG)},
  volume={25},
  number={3},
  pages={1214--1220},
  year={2006},
  publisher={ACM New York, NY, USA}
}

@incollection{kanamori2012region,
  title={Region matching with proxy ellipses for coloring hand-drawn animations},
  author={Kanamori, Yoshihiro},
  booktitle={SIGGRAPH Asia 2012 Technical Briefs},
  pages={1--4},
  year={2012}
}

@incollection{sato2014reference,
  title={Reference-based manga colorization by graph correspondence using quadratic programming},
  author={Sato, Kazuhiro and Matsui, Yusuke and Yamasaki, Toshihiko and Aizawa, Kiyoharu},
  booktitle={SIGGRAPH Asia 2014 Technical Briefs},
  pages={1--4},
  year={2014}
}

@misc{pfnet_paintschainer,
  author = {{Preferred Networks, Inc., Taizan Yonetsuji}},
  title = {PaintsChainer},
  year = {2017},
  publisher = {GitHub},
  journal = {GitHub repository},
  howpublished = {\url{https://github.com/pfnet/PaintsChainer}}
}

@article{maejima2024continual,
  title={Continual few-shot patch-based learning for anime-style colorization},
  author={Maejima, Akinobu and Shinagawa, Seitaro and Kubo, Hiroyuki and Funatomi, Takuya and Yotsukura, Tatsuo and Nakamura, Satoshi and Mukaigawa, Yasuhiro},
  journal={Computational Visual Media},
  volume={10},
  number={4},
  pages={705--723},
  year={2024},
  publisher={Springer}
}

@article{cao2024animediffusion,
  title={AnimeDiffusion: Anime diffusion colorization},
  author={Cao, Yu and Meng, Xiangqiao and Mok, PY and Lee, Tong-Yee and Liu, Xueting and Li, Ping},
  journal={IEEE Transactions on Visualization and Computer Graphics},
  volume={30},
  number={10},
  pages={6956--6969},
  year={2024},
  publisher={IEEE}
}

@inproceedings{liu2025manganinja,
  title={Manganinja: Line art colorization with precise reference following},
  author={Liu, Zhiheng and Cheng, Ka Leong and Chen, Xi and Xiao, Jie and Ouyang, Hao and Zhu, Kai and Liu, Yu and Shen, Yujun and Chen, Qifeng and Luo, Ping},
  booktitle={Proceedings of the Computer Vision and Pattern Recognition Conference},
  pages={5666--5677},
  year={2025}
}

@inproceedings{zhang2021user,
  title={User-guided line art flat filling with split filling mechanism},
  author={Zhang, Lvmin and Li, Chengze and Simo-Serra, Edgar and Ji, Yi and Wong, Tien-Tsin and Liu, Chunping},
  booktitle={Proceedings of the IEEE/CVF conference on computer vision and pattern recognition},
  pages={9889--9898},
  year={2021}
}

@inproceedings{ci2018user,
  title={User-guided deep anime line art colorization with conditional adversarial networks},
  author={Ci, Yuanzheng and Ma, Xinzhu and Wang, Zhihui and Li, Haojie and Luo, Zhongxuan},
  booktitle={Proceedings of the 26th ACM international conference on Multimedia},
  pages={1536--1544},
  year={2018}
}

@article{zhang2018two,
  title={Two-stage sketch colorization},
  author={Zhang, Lvmin and Li, Chengze and Wong, Tien-Tsin and Ji, Yi and Liu, Chunping},
  journal={ACM Transactions on Graphics (TOG)},
  volume={37},
  number={6},
  pages={1--14},
  year={2018},
  publisher={ACM New York, NY, USA}
}

@article{zou2019language,
  title={Language-based colorization of scene sketches},
  author={Zou, Changqing and Mo, Haoran and Gao, Chengying and Du, Ruofei and Fu, Hongbo},
  journal={ACM Transactions on Graphics (TOG)},
  volume={38},
  number={6},
  pages={1--16},
  year={2019},
  publisher={ACM New York, NY, USA}
}

@inproceedings{ishii2020confidence,
  title={Confidence-aware practical anime-style colorization},
  author={Ishii, Daichi and Kubo, Hiroyuki and Shinagawa, Seitaro and Maejima, Akinobu and Funatomi, Takuya and Nakamura, Satoshi and Mukaigawa, Yasuhiro},
  booktitle={Special Interest Group on Computer Graphics and Interactive Techniques Conference Talks},
  pages={1--2},
  year={2020}
}

@inproceedings{yan2025image,
  title={Image Referenced Sketch Colorization Based on Animation Creation Workflow},
  author={Yan, Dingkun and Wang, Xinrui and Li, Zhuoru and Saito, Suguru and Iwasawa, Yusuke and Matsuo, Yutaka and Guo, Jiaxian},
  booktitle={Proceedings of the Computer Vision and Pattern Recognition Conference},
  pages={23391--23400},
  year={2025}
}

@inproceedings{akita2020colorization,
  title={Colorization of line drawings with empty pupils},
  author={Akita, Kenta and Morimoto, Yuki and Tsuruno, Reiji},
  booktitle={Computer Graphics Forum},
  volume={39},
  number={7},
  pages={601--610},
  year={2020},
  organization={Wiley Online Library}
}

@inproceedings{dai2024learning,
  title={Learning inclusion matching for animation paint bucket colorization},
  author={Dai, Yuekun and Zhou, Shangchen and Li, Qinyue and Li, Chongyi and Loy, Chen Change},
  booktitle={Proceedings of the IEEE/CVF Conference on Computer Vision and Pattern Recognition},
  pages={25544--25553},
  year={2024}
}

@inproceedings{zhang2017style,
  title={Style transfer for anime sketches with enhanced residual u-net and auxiliary classifier gan},
  author={Zhang, Lvmin and Ji, Yi and Lin, Xin and Liu, Chunping},
  booktitle={2017 4th IAPR Asian conference on pattern recognition (ACPR)},
  pages={506--511},
  year={2017},
  organization={IEEE}
}

@inproceedings{verduyn2024towards,
  title={Towards Flat Color Prediction for Comics},
  author={Verduyn, Marnix and Tuytelaars, Tinne and others},
  booktitle={AI for Visual Arts Workshop at the European Conference on Computer Vision 2024, Date: 2024/09/29-2024/10/04, Location: Milano, Italy},
  year={2024}
}

@inproceedings{fourey2018fast,
  title={A fast and efficient semi-guided algorithm for flat coloring line-arts},
  author={Fourey, S{\'e}bastien and Tschumperl{\'e}, David and Revoy, David},
  booktitle={International Symposium on Vision, Modeling and Visualization},
  year={2018}
}

@article{zhang2009vectorizing,
  title={Vectorizing cartoon animations},
  author={Zhang, Song-Hai and Chen, Tao and Zhang, Yi-Fei and Hu, Shi-Min and Martin, Ralph R},
  journal={IEEE Transactions on Visualization and Computer Graphics},
  volume={15},
  number={4},
  pages={618--629},
  year={2009},
  publisher={IEEE}
}

@inproceedings{roy2019automation,
  title={Automation accuracy is good, but high controllability may be better},
  author={Roy, Quentin and Zhang, Futian and Vogel, Daniel},
  booktitle={Proceedings of the 2019 CHI Conference on Human Factors in Computing Systems},
  pages={1--8},
  year={2019}
}

@inproceedings{oh2018lead,
  title={I lead, you help but only with enough details: Understanding user experience of co-creation with artificial intelligence},
  author={Oh, Changhoon and Song, Jungwoo and Choi, Jinhan and Kim, Seonghyeon and Lee, Sungwoo and Suh, Bongwon},
  booktitle={Proceedings of the 2018 CHI conference on human factors in computing systems},
  pages={1--13},
  year={2018}
}

@inproceedings{fan2019collabdraw,
  title={Collabdraw: an environment for collaborative sketching with an artificial agent},
  author={Fan, Judith E and Dinculescu, Monica and Ha, David},
  booktitle={Proceedings of the 2019 Conference on Creativity and Cognition},
  pages={556--561},
  year={2019}
}

@inproceedings{rezwana2022understanding,
  title={Understanding user perceptions, collaborative experience and user engagement in different human-AI interaction designs for co-creative systems},
  author={Rezwana, Jeba and Maher, Mary Lou},
  booktitle={Proceedings of the 14th Conference on Creativity and Cognition},
  pages={38--48},
  year={2022}
}

@inproceedings{bird2024artists,
  title={Artists and AI: Creative Interactions and Tensions},
  author={Bird, Charlotte},
  booktitle={Extended Abstracts of the CHI Conference on Human Factors in Computing Systems},
  pages={1--6},
  year={2024}
}

@inproceedings{liapis2022need,
  title={The Need for Explainability in AI-Based Creativity Support Tools},
  author={Liapis, Antonios and Zhu, Jichen},
  booktitle={Proceedings of the Human Centered AI workshop at NeurIPS 2022},
  year={2022}
}

@article{pei2024human,
  title={Human--AI Co-Drawing: Studying Creative Efficacy and Eye Tracking in Observation and Cooperation},
  author={Pei, Yuying and Wang, Linlin and Xue, Chengqi},
  journal={Applied Sciences},
  volume={14},
  number={18},
  pages={8203},
  year={2024},
  publisher={MDPI}
}

@article{xu2023everyone,
  title={Is everyone an artist? A study on user experience of AI-based painting system},
  author={Xu, Junping and Zhang, Xiaolin and Li, Hui and Yoo, Chaemoon and Pan, Younghwan},
  journal={Applied Sciences},
  volume={13},
  number={11},
  pages={6496},
  year={2023},
  publisher={MDPI}
}

@inproceedings{sykora2009lazybrush,
  title={Lazybrush: Flexible painting tool for hand-drawn cartoons},
  author={S{\`y}kora, Daniel and Dingliana, John and Collins, Steven},
  booktitle={Computer Graphics Forum},
  volume={28},
  number={2},
  pages={599--608},
  year={2009},
  organization={Wiley Online Library}
}

@inproceedings{huang2023anifacedrawing,
  title={Anifacedrawing: Anime portrait exploration during your sketching},
  author={Huang, Zhengyu and Xie, Haoran and Fukusato, Tsukasa and Miyata, Kazunori},
  booktitle={ACM SIGGRAPH 2023 conference proceedings},
  pages={1--11},
  year={2023}
}

@inproceedings{kim2022colorbo,
  title={Colorbo: Envisioned mandala coloringthrough human-ai collaboration},
  author={Kim, Eunseo and Hong, Jeongmin and Lee, Hyuna and Ko, Minsam},
  booktitle={Proceedings of the 27th International Conference on Intelligent User Interfaces},
  pages={15--26},
  year={2022}
}

@inproceedings{yan2022flatmagic,
  title={FlatMagic: Improving flat colorization through AI-driven design for digital comic professionals},
  author={Yan, Chuan and Chung, John Joon Young and Kiheon, Yoon and Gingold, Yotam and Adar, Eytan and Hong, Sungsoo Ray},
  booktitle={Proceedings of the 2022 CHI conference on human factors in computing systems},
  pages={1--17},
  year={2022}
}

@article{bao2019scribble,
  title={Scribble-based colorization for creating smooth-shaded vector graphics},
  author={Bao, Bin and Fu, Hongbo},
  journal={Computers \& Graphics},
  volume={81},
  pages={73--81},
  year={2019},
  publisher={Elsevier}
}

@inproceedings{benedetti2014painting,
  title={Painting with Bob: assisted creativity for novices},
  author={Benedetti, Luca and Winnem{\"o}ller, Holger and Corsini, Massimiliano and Scopigno, Roberto},
  booktitle={Proceedings of the 27th annual ACM symposium on User interface software and technology},
  pages={419--428},
  year={2014}
}

@inproceedings{jalal2015color,
  title={Color portraits: From color picking to interacting with color},
  author={Jalal, Ghita and Maudet, Nolwenn and Mackay, Wendy E},
  booktitle={Proceedings of the 33rd Annual ACM Conference on Human Factors in Computing Systems},
  pages={4207--4216},
  year={2015}
}

@article{zhang2023image,
  title={Image inpainting based on deep learning: A review},
  author={Zhang, Xiaobo and Zhai, Donghai and Li, Tianrui and Zhou, Yuxin and Lin, Yang},
  journal={Information Fusion},
  volume={90},
  pages={74--94},
  year={2023},
  publisher={Elsevier}
}

@inproceedings{ronneberger2015u,
  title={U-net: Convolutional networks for biomedical image segmentation},
  author={Ronneberger, Olaf and Fischer, Philipp and Brox, Thomas},
  booktitle={International Conference on Medical image computing and computer-assisted intervention},
  pages={234--241},
  year={2015},
  organization={Springer}
}

@inproceedings{vogel2007shift,
  title={Shift: a technique for operating pen-based interfaces using touch},
  author={Vogel, Daniel and Baudisch, Patrick},
  booktitle={Proceedings of the SIGCHI conference on Human factors in computing systems},
  pages={657--666},
  year={2007}
}

@article{xing2024tooncrafter,
  title={Tooncrafter: Generative cartoon interpolation},
  author={Xing, Jinbo and Liu, Hanyuan and Xia, Menghan and Zhang, Yong and Wang, Xintao and Shan, Ying and Wong, Tien-Tsin},
  journal={ACM Transactions on Graphics (TOG)},
  volume={43},
  number={6},
  pages={1--11},
  year={2024},
  publisher={ACM New York, NY, USA}
}

@inproceedings{chilana2015user,
  title={From user-centered to adoption-centered design: a case study of an HCI research innovation becoming a product},
  author={Chilana, Parmit K and Ko, Amy J and Wobbrock, Jacob},
  booktitle={Proceedings of the 33rd Annual ACM Conference on Human Factors in Computing Systems},
  pages={1749--1758},
  year={2015}
}

@article{likert1932technique,
  title={A technique for the measurement of attitudes.},
  author={Likert, Rensis},
  journal={Archives of psychology},
  year={1932}
}

@inproceedings{bevan2015iso,
  title={ISO 9241-11 revised: What have we learnt about usability since 1998?},
  author={Bevan, Nigel and Carter, James and Harker, Susan},
  booktitle={International conference on human-computer interaction},
  pages={143--151},
  year={2015},
  organization={Springer}
}

@article{woolson2007wilcoxon,
  title={Wilcoxon signed-rank test},
  author={Woolson, Robert F},
  journal={Wiley encyclopedia of clinical trials},
  pages={1--3},
  year={2007},
  publisher={Wiley Online Library}
}

@book{tukey1977exploratory,
  title={Exploratory data analysis},
  author={Tukey, John Wilder and others},
  volume={2},
  year={1977},
  publisher={Springer}
}

@inproceedings{baudisch1998don,
  title={Don’t Click--Paint! Applying the Painting Metaphor to Query Interfaces and Personalization},
  author={Baudisch, Patrick},
  booktitle={Proceedings of UIST'98},
  pages={65--66},
  year={1998}
}

@misc{CLIPSTUDIO,
  title        = {CLIP STUDIO PAINT},
  author       = {Celsys},
  year         = 2025,
  note         = {Retrieved September 11, 2025 from \url{https://www.clipstudio.net}},
  howpublished = {Website}
}

@misc{AnimeIndustryReport,
  title        = {Anime Industry Report 2024 Summary\_20250321},
  author       = {The Association of Japanese Animation},
  year         = 2025,
    note = {Retrieved September 11, 2025 from https://aja.gr.jp/download/anime-industry-report-2024-summary\_20250321},
  howpublished = {Website}
}

@misc{MyAnimeList,
  title        = {MyAnimeList},
  author       = {MyAnimeList Co.,Ltd.},
  year         = 2025,
  note         = {Retrieved September 11, 2025 from \url{https://myanimelist.net}},
  howpublished = {Website}
}

@misc{BroadcastStats,
  title        = {{Ni-kuru mono ga fueta tte honto ka? (36) 2024-nen kakuhoban [Has Two-Cour Anime Really Increased? (36): Final Report for 2024]}},
  author       = {{Anime Chosashitsu (Kari) [Anime Research Lab (provisional)]}},
  year         = 2025,
  note         = {Retrieved September 11, 2025 from \url{http://anime-research.seesaa.net/article/516525382.html} in Japanese},
  howpublished = {Website}
}

@book{lazar2017research,
  title={Research methods in human-computer interaction},
  author={Lazar, Jonathan and Feng, Jinjuan Heidi and Hochheiser, Harry},
  year={2017},
  publisher={Morgan Kaufmann}
}

@article{hurlbert1984pseudoreplication,
  title={Pseudoreplication and the design of ecological field experiments},
  author={Hurlbert, Stuart H},
  journal={Ecological monographs},
  volume={54},
  number={2},
  pages={187--211},
  year={1984},
  publisher={Wiley Online Library}
}

@article{singh2025systematic,
  title={A Systematic Review of Human-AI Co-Creativity},
  author={Singh, Saloni and Hindriks, Koen and Heylen, Dirk and Baraka, Kim},
  journal={arXiv preprint arXiv:2506.21333},
  year={2025}
}

@article{centeio2023exploring,
  title={Exploring the effect of automation failure on the human’s trustworthiness in human-agent teamwork},
  author={Centeio Jorge, Carolina and Bouman, Nikki H and Jonker, Catholijn M and Tielman, Myrthe L},
  journal={Frontiers in Robotics and AI},
  volume={10},
  pages={1143723},
  year={2023},
  publisher={Frontiers Media SA}
}

@article{madhavan2006automation,
  title={Automation failures on tasks easily performed by operators undermine trust in automated aids},
  author={Madhavan, Poornima and Wiegmann, Douglas A and Lacson, Frank C},
  journal={Human factors},
  volume={48},
  number={2},
  pages={241--256},
  year={2006},
  publisher={SAGE Publications Sage CA: Los Angeles, CA}
}

@article{rezwana2023designing,
  title={Designing creative AI partners with COFI: A framework for modeling interaction in human-AI co-creative systems},
  author={Rezwana, Jeba and Maher, Mary Lou},
  journal={ACM Transactions on Computer-Human Interaction},
  volume={30},
  number={5},
  pages={1--28},
  year={2023},
  publisher={ACM New York, NY}
}

@inproceedings{rezwana2021cofi,
  title={COFI: A Framework for Modeling Interaction in Human-AI Co-Creative Systems.},
  author={Rezwana, Jeba and Maher, Mary Lou},
  booktitle={ICCC},
  pages={444--448},
  year={2021}
}

@article{wang2009mean,
  title={Mean squared error: Love it or leave it? A new look at signal fidelity measures},
  author={Wang, Zhou and Bovik, Alan C},
  journal={IEEE signal processing magazine},
  volume={26},
  number={1},
  pages={98--117},
  year={2009},
  publisher={IEEE}
}

@inproceedings{zhang2018unreasonable,
  title={The unreasonable effectiveness of deep features as a perceptual metric},
  author={Zhang, Richard and Isola, Phillip and Efros, Alexei A and Shechtman, Eli and Wang, Oliver},
  booktitle={Proceedings of the IEEE conference on computer vision and pattern recognition},
  pages={586--595},
  year={2018}
}

@article{rieger2022human,
  title={Human performance consequences of automated decision aids: The impact of time pressure},
  author={Rieger, Tobias and Manzey, Dietrich},
  journal={Human factors},
  volume={64},
  number={4},
  pages={617--634},
  year={2022},
  publisher={Sage Publications Sage CA: Los Angeles, CA}
}

@article{vimpari2023adapt,
  title={“An adapt-or-die type of situation”: perception, adoption, and use of text-to-image-generation AI by game industry professionals},
  author={Vimpari, Veera and Kultima, Annakaisa and H{\"a}m{\"a}l{\"a}inen, Perttu and Guckelsberger, Christian},
  journal={Proceedings of the ACM on Human-Computer Interaction},
  volume={7},
  number={CHI PLAY},
  pages={131--164},
  year={2023},
  publisher={ACM New York, NY, USA}
}

@inproceedings{inie2023designing,
  title={Designing participatory ai: Creative professionals’ worries and expectations about generative ai},
  author={Inie, Nanna and Falk, Jeanette and Tanimoto, Steve},
  booktitle={Extended Abstracts of the 2023 CHI Conference on Human Factors in Computing Systems},
  pages={1--8},
  year={2023}
}

@inproceedings{palani2024evolving,
  title={Evolving roles and workflows of creative practitioners in the age of generative AI},
  author={Palani, Srishti and Ramos, Gonzalo},
  booktitle={Proceedings of the 16th Conference on Creativity \& Cognition},
  pages={170--184},
  year={2024}
}

@article{ogawa2025understanding,
author = {Ogawa, Nami and Okafuji, Yuki and Hatada, Yuji and Baba, Jun},
title = {Understanding Collaboration between Professional Designers and Decision-making AI: A Case Study in the Workplace},
year = {2025},
issue_date = {November 2025},
publisher = {Association for Computing Machinery},
address = {New York, NY, USA},
volume = {9},
number = {7},
url = {https://doi.org/10.1145/3757686},
doi = {10.1145/3757686},
journal = {Proc. ACM Hum.-Comput. Interact.},
month = oct,
articleno = {CSCW505},
numpages = {26}
}

@inproceedings{uusitalo2024clay,
  title={” Clay to Play With”: Generative AI Tools in UX and Industrial Design Practice},
  author={Uusitalo, Severi and Salovaara, Antti and Jokela, Tero and Salmimaa, Marja},
  booktitle={Proceedings of the 2024 ACM Designing Interactive Systems Conference},
  pages={1566--1578},
  year={2024}
}

@inproceedings{frich2019mapping,
  title={Mapping the landscape of creativity support tools in HCI},
  author={Frich, Jonas and MacDonald Vermeulen, Lindsay and Remy, Christian and Biskjaer, Michael Mose and Dalsgaard, Peter},
  booktitle={Proceedings of the 2019 CHI conference on human factors in computing systems},
  pages={1--18},
  year={2019}
}

@article{tang2025generative,
  title={Generative ai for cel-animation: A survey},
  author={Tang, Yunlong and Guo, Junjia and Liu, Pinxin and Wang, Zhiyuan and Hua, Hang and Zhong, Jia-Xing and Xiao, Yunzhong and Huang, Chao and Song, Luchuan and Liang, Susan and others},
  journal={arXiv preprint arXiv:2501.06250},
  year={2025}
}

@misc{rai2025sketchanimationstateoftheartreport,
      title={Sketch Animation: State-of-the-art Report}, 
      author={Gaurav Rai and Ojaswa Sharma},
      year={2025},
      eprint={2510.10218},
      archivePrefix={arXiv},
      primaryClass={cs.GR},
      url={https://arxiv.org/abs/2510.10218}, 
}

@article{dong2025kisscolor,
  title={KISSColor: Kinetic and Intuitive Stroke Stretching for Vector Drawing Colorization},
  author={Dong, Yiming and Xin, Hongxu and Dou, Zhiyang and Xu, Rui and Liu, Yuan and Chen, Shuangmin and Xin, Shiqing and Tu, Changhe and Komura, Taku and Wang, Wenping},
  journal={ACM Transactions on Graphics (TOG)},
  volume={44},
  number={6},
  pages={1--13},
  year={2025},
  publisher={ACM New York, NY, USA}
}

@incollection{allen2024fast,
  title={Fast Leak-Resistant Segmentation for Anime Line Art},
  author={Allen, Benjamin and Maejima, Akinobu and Anjyo, Ken},
  booktitle={SIGGRAPH Asia 2024 Technical Communications},
  pages={1--4},
  year={2024}
}

@incollection{frich2018hci,
  title={Why HCI and creativity research must collaborate to develop new creativity support tools},
  author={Frich, Jonas and Biskjaer, Michael Mose and Dalsgaard, Peter},
  booktitle={Proceedings of the Technology, Mind, and Society},
  pages={1--6},
  year={2018}
}

@article{rhodes1961analysis,
  title={An analysis of creativity},
  author={Rhodes, Mel},
  journal={The Phi delta kappan},
  volume={42},
  number={7},
  pages={305--310},
  year={1961},
  publisher={JSTOR}
}

@inproceedings{swaroop2024accuracy,
  title={Accuracy-time tradeoffs in AI-assisted decision making under time pressure},
  author={Swaroop, Siddharth and Bu{\c{c}}inca, Zana and Gajos, Krzysztof Z and Doshi-Velez, Finale},
  booktitle={Proceedings of the 29th International Conference on Intelligent User Interfaces},
  pages={138--154},
  year={2024}
}

@article{cao2023time,
  title={How time pressure in different phases of Decision-Making influences Human-AI collaboration},
  author={Cao, Shiye and Gomez, Catalina and Huang, Chien-Ming},
  journal={Proceedings of the ACM on Human-computer Interaction},
  volume={7},
  number={CSCW2},
  pages={1--26},
  year={2023},
  publisher={ACM New York, NY, USA}
}

@inproceedings{remy2020evaluating,
  title={Evaluating creativity support tools in HCI research},
  author={Remy, Christian and MacDonald Vermeulen, Lindsay and Frich, Jonas and Biskjaer, Michael Mose and Dalsgaard, Peter},
  booktitle={Proceedings of the 2020 ACM designing interactive systems conference},
  pages={457--476},
  year={2020}
}

@article{wang2025animagents,
  title={AnimAgents: Coordinating Multi-Stage Animation Pre-Production with Human-Multi-Agent Collaboration},
  author={Wang, Wen-Fan and Lu, Chien-Ting and Ng, Jin Ping and Chiu, Yi-Ting and Lee, Ting-Ying and Wang, Miaosen and Chen, Bing-Yu and Chen, Xiang'Anthony'},
  journal={arXiv preprint arXiv:2511.17906},
  year={2025}
}

@article{westphal2023decision,
  title={Decision control and explanations in human-AI collaboration: Improving user perceptions and compliance},
  author={Westphal, Monika and V{\"o}ssing, Michael and Satzger, Gerhard and Yom-Tov, Galit B and Rafaeli, Anat},
  journal={Computers in Human Behavior},
  volume={144},
  pages={107714},
  year={2023},
  publisher={Elsevier}
}

@inproceedings{kocielnik2019will,
  title={Will you accept an imperfect ai? exploring designs for adjusting end-user expectations of ai systems},
  author={Kocielnik, Rafal and Amershi, Saleema and Bennett, Paul N},
  booktitle={Proceedings of the 2019 CHI conference on human factors in computing systems},
  pages={1--14},
  year={2019}
}

@inproceedings{yin2019understanding,
  title={Understanding the effect of accuracy on trust in machine learning models},
  author={Yin, Ming and Wortman Vaughan, Jennifer and Wallach, Hanna},
  booktitle={Proceedings of the 2019 chi conference on human factors in computing systems},
  pages={1--12},
  year={2019}
}

@article{dietvorst2015algorithm,
  title={Algorithm aversion: people erroneously avoid algorithms after seeing them err.},
  author={Dietvorst, Berkeley J and Simmons, Joseph P and Massey, Cade},
  journal={Journal of experimental psychology: General},
  volume={144},
  number={1},
  pages={114},
  year={2015},
  publisher={American Psychological Association}
}

@article{shneiderman2007creativity,
  title={Creativity support tools: accelerating discovery and innovation},
  author={Shneiderman, Ben},
  journal={Communications of the ACM},
  volume={50},
  number={12},
  pages={20--32},
  year={2007},
  publisher={ACM New York, NY, USA}
}

@inproceedings{chung2022artist,
  title={Artist support networks: Implications for future creativity support tools},
  author={Chung, John Joon Young and He, Shiqing and Adar, Eytan},
  booktitle={Proceedings of the 2022 ACM Designing Interactive Systems Conference},
  pages={232--246},
  year={2022}
}

@inproceedings{frich2019strategies,
  title={Strategies in Creative Professionals' Use of Digital Tools Across Domains},
  author={Frich, Jonas and Biskjaer, Michael Mose and MacDonald Vermeulen, Lindsay and Remy, Christian and Dalsgaard, Peter},
  booktitle={Proceedings of the 2019 Conference on Creativity and Cognition},
  pages={210--221},
  year={2019}
}

@article{grassini2025artificial,
  title={Artificial creativity? Evaluating AI against human performance in creative interpretation of visual stimuli},
  author={Grassini, Simone and Koivisto, Mika},
  journal={International journal of human--computer interaction},
  volume={41},
  number={7},
  pages={4037--4048},
  year={2025},
  publisher={Taylor \& Francis}
}

@article{mcgrath2025collaborative,
  title={Collaborative human-AI trust (CHAI-T): A process framework for active management of trust in human-AI collaboration},
  author={McGrath, Melanie J and Duenser, Andreas and Lacey, Justine and Paris, Cecile},
  journal={Computers in Human Behavior: Artificial Humans},
  pages={100200},
  year={2025},
  publisher={Elsevier}
}

@inproceedings{gmeiner2023exploring,
  title={Exploring challenges and opportunities to support designers in learning to co-create with AI-based manufacturing design tools},
  author={Gmeiner, Frederic and Yang, Humphrey and Yao, Lining and Holstein, Kenneth and Martelaro, Nikolas},
  booktitle={Proceedings of the 2023 CHI Conference on Human Factors in Computing Systems},
  pages={1--20},
  year={2023}
}

@misc{popova2023co,
  title={Co-creating Futures for Integrating Generative AI into the Designers’ Workflow},
  author={Popova, Victoria},
  year={2023}
}

\appendix
\appendix

\section{Analyses of Enclosed Regions in Anime Images}
\label{appendix:region_analysis}

\begin{figure*}[t]
  \centering
  \includegraphics[width=\linewidth]{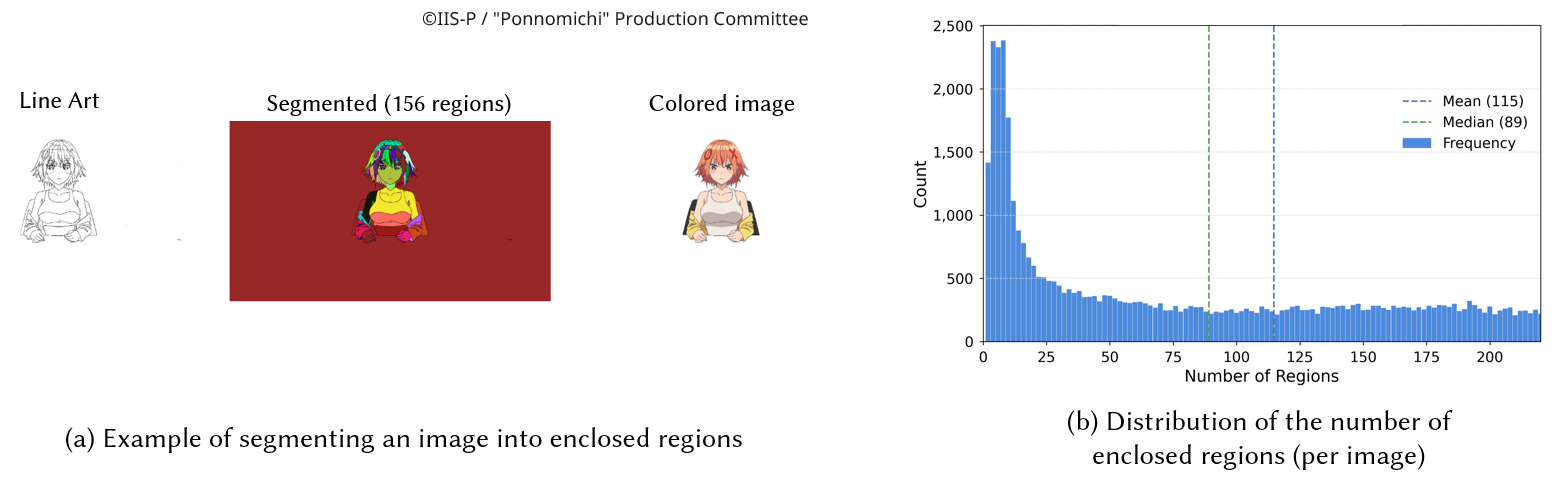}
  \caption[Number of regions]{
(a) Example of segmenting an image into enclosed regions and (b) the distribution of the number of enclosed regions (per image) across all frames of an anime.
  }
  \Description{The figure consists of two panels illustrating the complexity of anime image segmentation.
Panel (a) shows a visual example of an anime character in three stages: raw line art, a segmented visualization where 156 distinct enclosed regions are filled with random colors to highlight complexity, and the final colored image.
Panel (b) is a histogram showing the distribution of the number of enclosed regions per image across a dataset. The x-axis represents the Number of Regions (ranging from 0 to over 200), and the y-axis represents the Count. The distribution is right-skewed. Vertical dashed lines indicate that the Median number of regions is 89 and the Mean is 115.}
  \label{fig:num_regions}
\end{figure*}

\begin{figure*}[t]
  \centering
  \includegraphics[width=\linewidth]{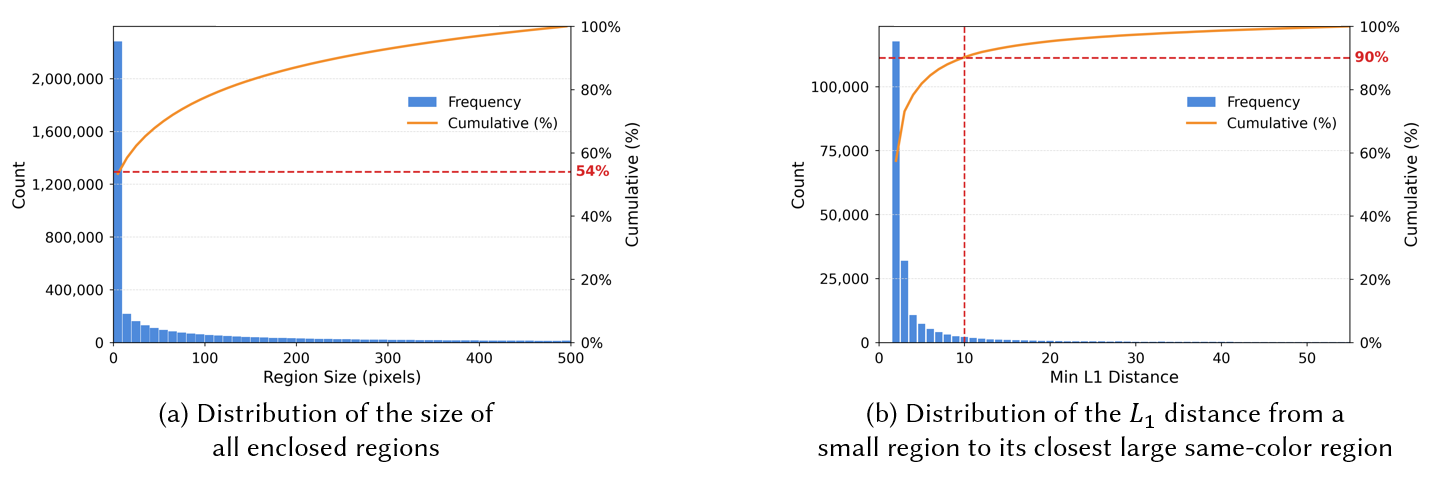}
\caption[Region and gap analyses]{(a) The size distribution of all enclosed regions, and
(b) the distribution of the $L_1$ distance from regions smaller than $10$ pixels to their closest large region of the same color.}
\Description{The figure comprises two histograms with overlaying cumulative distribution curves (orange lines).
Panel (a) displays the region size in pixels on the x-axis (0 to 500) and frequency on the y-axis. The distribution is extremely right-skewed, with a massive spike at the smallest size. A horizontal red dashed line highlights that $54\%$ of all enclosed regions are extremely small (located in the first bin, representing under 10 pixels).
Panel (b) displays the Minimum $L_1$ distance on the x-axis (0 to 50) and frequency on the y-axis. It shows the proximity of small regions to their closest large same-color region. This is also right-skewed. Intersecting red dashed lines indicate that $90\%$ of small regions are located within an $L_1$ distance of $10$ pixels from a larger region of the same color.}
  \label{fig:region2}
\end{figure*}

To investigate the characteristics of enclosed regions in anime-style images, we performed a flood-fill-based (BFS) segmentation over all frames in our dataset ($51,892$ frames from $12$ episodes of a professionally produced anime series). We analyzed three aspects: the number of regions, their size distribution, and the distance of small regions to large regions of the same color.

\subsection{Number of Enclosed Regions}
\cref{fig:num_regions}a shows an example of the segmentation process, while~\cref{fig:num_regions}b presents the distribution of the number of enclosed regions across all frames. On average, each frame contained $115$ enclosed regions, with a median of $89$. This indicates that anime images typically include a large number of enclosed areas.

\subsection{Region Size Distribution}
We next examined the distribution of region sizes, measured in pixels. \cref{fig:region2}a shows the frequency of each size. We observed that regions of size $10$ pixels or smaller occur in sufficient numbers (approximately $54\%$), which motivated our initial definition of ``small regions'' as those with under $10$ pixels when considering potential unpainted gaps.

\subsection{Distance to Large Same-Color Regions}
Finally, we investigated how small regions (size $\leq 10$) relate to larger regions of the same color. We computed the $L_1$ distance between each small region and its nearest large same-color region, based on the closest pair of pixels between them (\cref{fig:region2}b). The cumulative distribution reveals that the vast majority of small regions (approximately $90\%$) have a same-color large region within a distance of $10$ pixels. This confirms the tendency observed in the formative study that small enclosed regions often share the same color with their spatially neighboring areas.

\section{Supplementary Statistical Analysis}
\label{appendix:sub_stat}

\begin{figure*}[t]
  \centering
  \includegraphics[width=\linewidth]{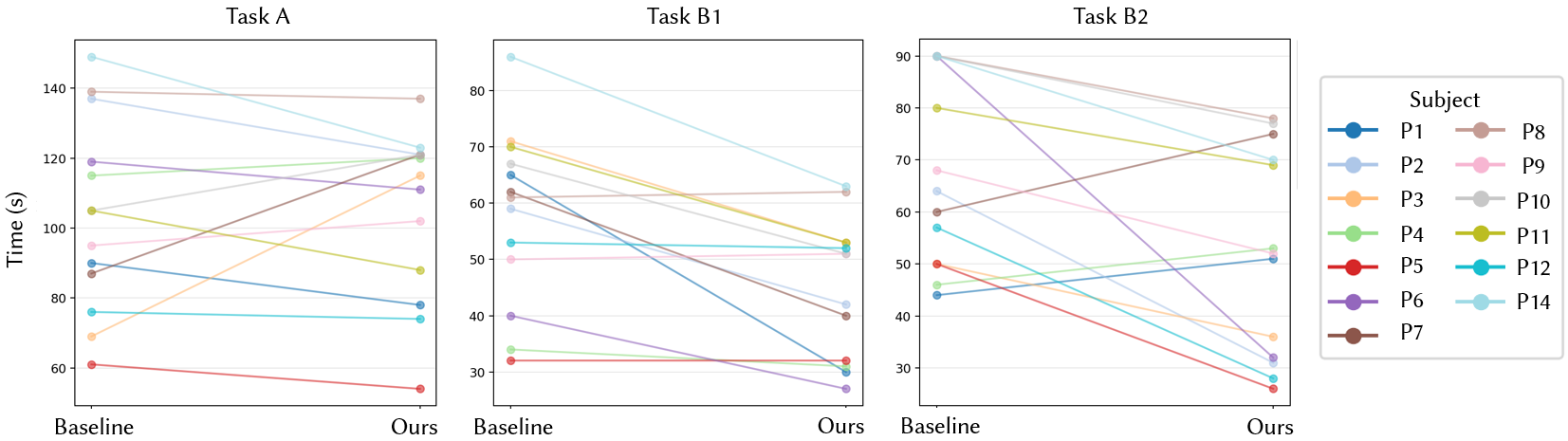}
    \caption[Raw data]{Visualization of paired differences for each participant.}
\Description{The figure presents three slope graphs, one for each task (Task A, Task B1, and Task B2), visualizing the change in completion time (seconds) for 13 individual participants (P1–P14).
The x-axis in each graph connects the "Baseline" condition on the left to "Ours" on the right.
- Task A (left panel): The slopes are mixed, with lines going both up and down, indicating that time differences varied by participant.
- Task B1 (center) and Task B2 (right): These panels show a consistent downward trend. Most lines slope down from Baseline to Ours, visually demonstrating that the majority of participants completed these tasks faster using the proposed method.}
  \label{fig:time_vary}
\end{figure*}

We conducted a one-sided paired $t$-test as a supplementary analysis to reinforce the primary findings for the user study. \cref{fig:time_vary} visualizes the paired differences for each participant.

For Task A, the mean difference was $-2.33$ ($SD = 16.26$, $99\%$ CI $[-16.91, 12.24]$), and the $t$-test showed no significant difference ($t(11) = -0.497, p = .31440$).
For Task B1, the mean difference was $-12.54$ ($SD = 11.30$, $99\%$ CI $[-22.11, -2.96]$), showing a significant reduction in time ($t(12) = -3.999, p = .00088$).
For Task B2, the mean difference was $-14.36$ ($SD = 12.73$, $99\%$ CI $[-26.53, -2.20]$), also indicating a significant reduction ($t(10) = -3.742, p = .00192$).

\end{document}